# Earthquake Nowcasting with Deep Learning

Geoffrey Fox[1], John Rundle[2], Andrea Donnellan[3], Bo Feng[4],

1) Biocomplexity Institute and Initiative, University of Virginia, Charlottesville, VA, 22904, USA; Computer Science Dept., University of Virginia, Charlottesville, VA 22904, USA
2) University of California, 95616, Davis, CA, USA; Jet Propulsion Laboratory, California Institute of Technology, 91109, Pasadena, CA, USA; Santa Fe Institute, 87501, Santa Fe, NM, USA; Jet Propulsion Laboratory, California Institute of Technology, 91109, Pasadena, CA, USA; Santa Fe Institute, 87501, Santa Fe, NM, USA
3) Jet Propulsion Laboratory, California Institute of Technology, 91109, Pasadena, CA, USA
4) Indiana University, Bloomington, IN 47408, USA

## Abstract

We review previous approaches to nowcasting earthquakes and introduce new approaches based on deep learning using three distinct models based on recurrent neural networks and transformers. We discuss different choices for observables and measures presenting promising initial results for a region of Southern California from 1950-2020. Earthquake activity is predicted as a function of 0.1-degree spatial bins for time periods varying from two weeks to four years. The overall quality is measured by the Nash Sutcliffe Efficiency comparing the deviation of nowcast and observation with the variance over time in each spatial region. The software is available as open-source together with the preprocessed data from the USGS.

## 1 Introduction

Earthquake forecasting is an old and difficult problem with many interesting characteristics. In studying this we not only hope to shed light on this socioscientific challenge but also lead to new methods based on deep learning that can be applied in other areas. Perhaps the most important characteristic is the nature of its challenge. Namely, it is unlikely that building a new zettascale supercomputer will accurately simulate quakes and lead to reliable earthquake predictions [1]. As a phase transition in a system with unknown boundary conditions and phenomenological models (as for friction), it is not obvious that earthquakes are the solution of a set of differential equations or that accurate probabilities of large events can be computed. For these reasons, we have chosen recently to focus on earthquake nowcasting [2], which is the estimation of risk in the present, the immediate past, and the near future.

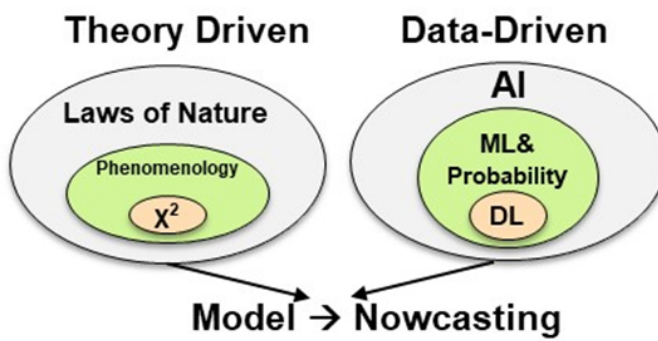

Fig. 1: Comparison of traditional theory driven analysis (left) with the data-driven ideas

Rather, earthquake nowcasting is an archetype of data-intensive problems characteristic of the fourth paradigm of scientific discovery [3]; we suppose that the patterns of previous events hold clues to the future. This is a bit different from say other studies where machine learning has been successful; for example, language translation and image recognition are very complicated problems but one can see there are natural patterns from grammar, words, segments, colors, etc. that are clearly but complexly correlated with what we want to find out. Machine learning successfully learns this complex relationship between inputs and predictions.

Earthquake nowcasting challenges AI to discover "hidden variables" in a case where their existence is not as clear as in other successes of machine and deep learning. These ideas are illustrated in figure 1 where we are pursuing the right side data-intensive approach, rather than developing a theoretical model and determining unknown parameters from the data. In this paper, everything is hidden and determined by the data! Data-driven methods avoid the bias of incomplete theories but maybe there is not rich enough data to provide good insights. We partly address this here with a choice of "known inputs" which are natural mathematical expansion functions explained in Sec. 3.5.

The history of earthquakes is recorded as events detailing the (three-dimensional) position and size of each quake. These can be binned in space and time to generate geospatial time series corresponding to sequences in time labeled by spatial position (and bin size).  This class of problem is extremely common in both science (research) and commercial areas. Perhaps the most intense study of this problem has come from traffic transportation) studies with many innovations aimed at ride-hailing with papers from for example the companies Uber and Didi as well as academia. However, there are also many medical and earth/environmental science problems of this type  [4]–[9]. This problem class includes cases where there is a significant spatial structure that typically strongly affects nearby space points.

Details of the geometric structure are important, for example, in traffic-related problems where the road network and function of land use define a graph structure relating different spatial points [10]–[13]. The problem chosen is termed [14], [15] a spatial bag where there is spatial variation but it is not clearly linked to the geometric distance between spatial regions. Earthquakes have obvious coarse grain spatial structure but local structure comes from faults and we have not found this useful in the current analysis which therefore is in the spatial bag class. Hydrology study of catchments [4], [16] is also a spatial bag problem. Different catchments are related by values of static attributes such as aridity, elevation, and the nature of the soil and not directly by geometry.

In the following text, we first review the use of nowcasting and machine learning for earthquakes. Then we describe the data setup for deep learning and describe the three methods used in this work. We present nowcasting results from each of these and finish with a summary of some open questions in this promising but preliminary area. We hope that the deep learning approach could be more general and not require a priori guesswork as to what patterns were important.

## 1.1 Introduction to Earthquake Nowcasting

Earthquakes are a clear and present danger to world communities [17], and many earthquake forecasting methods have been proposed to date with little success.  A comprehensive recent review is given by [2].  In this paper, we continue the development of a new approach, earthquake nowcasting, using machine learning methods that have recently been developed.

Nowcasting is a term originating from economics and finance. It refers to the process of determining the uncertain state of the economy or markets at the *current time* by indirect means. By the current time we mean the present, the immediate past, and the very near future. Rundle and Donnellan [18], [19] and [2], [20]–[22] have applied this idea to seismically active regions, where the goal is to determine the current state of the fault system, and its current level of progress through the earthquake cycle. In our implementation of this idea, we use the global catalog of earthquakes, using "small" earthquakes to determine the level of hazard from "large" earthquakes in the region.

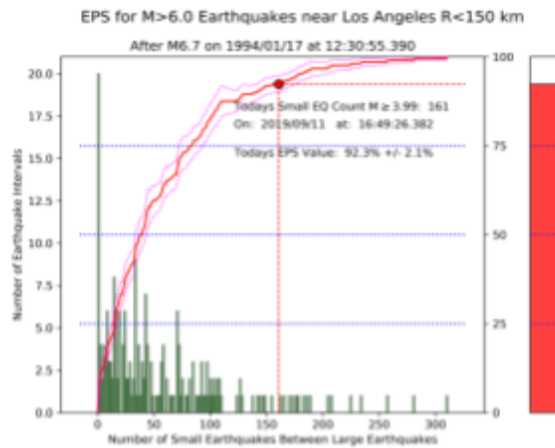

Fig. 2. Example of a nowcast for M6 earthquakes near Los Angeles

The original method is based on the idea of an earthquake cycle. A specific region and a specific large earthquake magnitude of interest were defined, ensuring that there is enough data to span at least ~20 or more large earthquake cycles in the region. An "Earthquake Potential Score" (EPS) was then defined as the cumulative probability distribution $P(n<n(t))$ for the current count n(t) for the small earthquakes in the region. Physically, the EPS corresponds to an estimate of the level of progress through the earthquake cycle in the defined region at the current time.

In the past, this determination of the state of a regional fault system has focused on trying to estimate the state of stress in the earth, its relation to the failure strength of the active faults in a region, and the rate of accumulation of tectonic stress. Determining the values of these parameters would allow researchers to estimate the proximity to failure of the faults in the region. This would be an answer to the question of "how far along is the region in the earthquake cycle?".

An example application of this method is shown in Figure 2. Here we show the EPS for a region surrounding Los Angeles within a circle of radius 150 Km, for earthquakes of magnitude M≥6 [21]–[23]. The last such earthquake was the Northridge, CA earthquake of January 17, 1994. The green vertical bars represent a histogram of the number of small earthquakes between large earthquakes M≥6 in a region 4000 KM x 4000 Km surrounding Los Angeles. The solid red line is the corresponding cumulative distribution function (CDF). The thin dashed lines represent the 68% confidence bound on the CDF. The red dot represents the number of small earthquakes that have occurred in the region since the Northridge event.
Radius of Gyration of Bursts

A different but related method uses small earthquake bursts that are strongly clustered in space and time ([18], [19]; Figure 3). These include seismic swarms and aftershock sequences. A readily observable property of these events, the radius of gyration ($R_G$), allows us to connect the bursts to the temporal occurrence of the largest *M*≥7 earthquakes in California since 1984.

In the Southern California earthquake catalog, we identify hundreds of these potentially coherent space-time structures in a region defined by a circle of radius 600 km around Los Angeles. We compute $R_G$ for each cluster, then filter them to identify those groups of events with large numbers of events closely clustered in space, which we call "compact" bursts. Our basic assumption is that these compact bursts reflect the dynamics associated with large earthquakes.

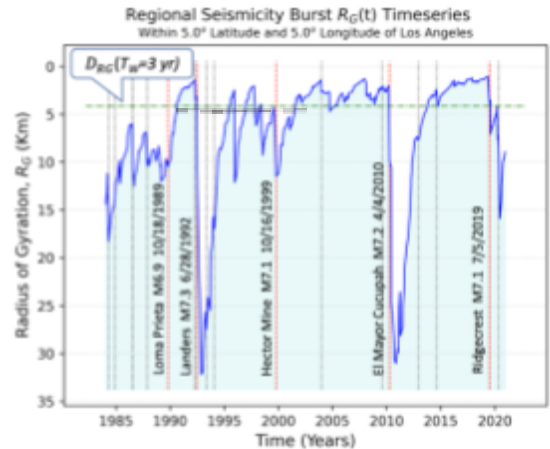

Fig. 3. Earthquake Burst Nowcasting

Once we have filtered the burst catalog, we apply an exponential moving average to construct a time series $R_G(t)$ for the Southern California region. We observe that the $R_G(t)$ of these bursts systematically decreases prior to large earthquakes, in a process that we might term "radial localization." The $R_G(t)$ then rapidly increases during an aftershock sequence, and a new cycle of "radial localization" then begins.

These time-series display cycles of recharge and discharge reminiscent of seismic stress accumulation and release in the elastic rebound process. This time series is shown in Figure 3. Vertical red dashed lines represent major earthquakes having magnitudes M≥7, vertical dotted lines represent earthquakes having 6≤M≤7. The green horizontal line near the top is a Decision Threshold $D_{RG}(T_W)$ used for an optimal Receiver Operating Characteristic analysis with an alert time of $T_W$ = 3 years [2], [20].

A similar time series that is based on different correlation metrics has been constructed by Rundle et al.[2], [20]. This method uses a new machine learning-based method for nowcasting earthquakes to image the time-dependent earthquake cycle, involving the classification of seismicity into characteristic patterns found by using Principal Component Analysis of regional seismicity ([24]–[26]). Patterns are found as eigenvectors of the cross-correlation matrix of a collection of seismicity time-series in a coarse-grained regional spatial grid (pattern recognition via unsupervised machine learning). Data from 1950-present was used, with magnitudes M≥3.29. The eigenvalues of this matrix represent the relative importance of the various eigenpatterns. Examples of the most prominent eigenvectors are shown in Rundle et al. [2], [20], [27].

We can now ask whether it might be possible to apply Machine Learning (ML) techniques to predict future values of the time series. The general method we have used is to divide the data into a training and a test set, the training period to fix the parameters, and the test set to evaluate the quality of the prediction. There are at least 3 techniques that we are at present in the process of evaluating. These are all one-step walk-forward methods in which a set of $N$ previous values $\{x_t, x_{t-1},..., x_{t-N}\}$ are used as the feature vector **x**. The label y that is assigned for the supervised learning task is the next value of the time series beyond $x_t$, y = $x_{t+1}$.

Once the model is trained, it is evaluated on the test or validation data. There are three methods that we are so far investigating based on:

- Random Forest Classifiers (RFC) [28]
- Auto Regressive Integrated Moving Average (ARIMA) [28]
- Long Short Term Memory Recurrent Neural Networks (LSTM RNN) [29] used in our deep learning work later
- Generative Adversarial Networks (GAN) [30]–[32]

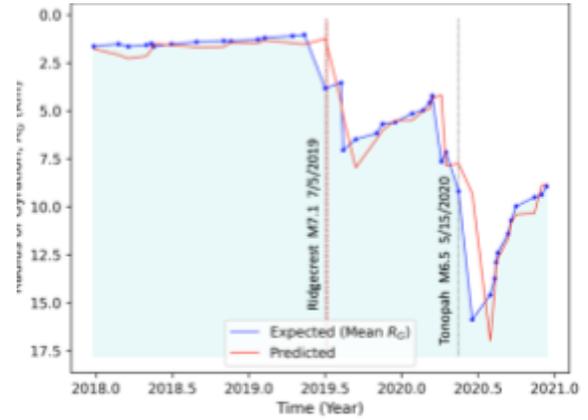

Briefly, the RFC [28] method uses an ensemble of decision trees to estimate model parameters using a classifier method based, for example, on the ID3 [33] or more recent classifier methods. Classification is carried out by optimizing the entropy or the Gini index. An example of the RFC method is shown here in Figure 4. It can be seen that the largest earthquakes are not predicted by this method and so we must search for other methods.

Fig 4; An example of a time series prediction using the RFC method.

### 1.2 Research objectives

The scientific objective is to improve the quality of Earthquake Nowcasting and we explore deep learning in a region of Southern California, similar to that used in previous work [20], [27]. This is quantified in a few metrics -- the Nash Sutcliffe Efficiency NSE [34] and in the visualization of predicted versus observed earthquake activity. NSE is a measure of a difference between the variation of the data over time with the discrepancy between observation and prediction so that predicting the observation to be the mean gives a default NSE of 0. The Nash Sutcliffe Efficiency NSE is given in equation (1) where quantity $Q^t_m$ is model prediction at time t for Quantity $Q$; $Q^t_o$ is the observed value of $Q$ at time t and $\overline{Q_o}$ is the mean value of $Q^t_o$ over time

$$NSE = 1 - \frac{\sum_{t=1}^{t=T}(Q^t_m - Q^t_o)^2}{\sum_{t=1}^{t=T}(Q^t_o - \overline{Q_o})^2} \quad (1)$$

We present results in the form of the normalized NSE, NNSE = 1/(2-NSE) which runs between 0 (bad) and 1 (perfection). This approach is common for the evaluation of time series models in other earth science fields [35], [36]. We supplement these overall averages with time-dependent plots of earthquake activity illustrated in sec. 5.

## 2 Data description
### 2.1 Quake Data

The earthquake data comes from USGS [37] and we have chosen 4 degrees of Latitude (32 to 36 N) and 6 degrees of Longitude (-120 to -114) region covering Southern California. The data runs from 1950 to the present day and is presented as events: magnitude, ground location, depth, and time. We have presented the data in time and space bins [38]. The time interval is daily but in our reference models, we accumulate this into fortnightly data. Southern California is divided into a 40 by 60 grid of 0.1 by 0.1-degree "pixels" which corresponds roughly to squares with an 11 km side, The dataset also includes an assignment of pixels to known faults and a list of the largest earthquakes in that region from 1950 until today.

We have chosen various samplings of the dataset to provide both input and predicted values. These include time ranges from a fortnight up to 4 years. Further, we calculate summed magnitudes and depths and counts of significant quakes (magnitude > 3.29). Other easily available quantities are powers of quake energy (using Energy ~ $10^{1.5m}$ where m is magnitude). We use the concept of "Energy averaging" when there are multiple events in a single space-time bin. So the magnitude $m_{bin}$ assigned to each bin is defined in equation 2 as "$log(Energy)$" where we sum over events of individual magnitudes $m_{event}$

$$m_{bin} = log(Energy) = \frac{1}{1.5}\log_{10} \sum_{in\ bin}^{Events} 10^{1.5m_{event}} \quad (2)$$

We also use energy averaging defined in equation (3) for quantities $Q$ such as the depth

$$Energy\ weighted\ Quantity\ Q_{bin} = \frac{\sum_{in\ bin}^{Events} 10^{1.5m_{event}} Q_{event}}{\sum_{in\ bin}^{Events} 10^{1.5m_{event}}} \quad (3)$$

The multiplicity per each bin is the number of events in a bin satisfying a criterion and are not "Energy averaged".

$$Multiplicity_{bin} = \sum_{in\ bin}^{Events} Multiplicity_{event} \quad (4)$$

We also looked at the powers of the energy $E^n$ instead of $m_{bin}$

$$E^n = \left(\sum_{in\ bin}^{Events} 10^{1.5m_{event}}\right)^n \quad \text{for n = 0.25, 0.5, 1} \quad (5)$$

As the exponent n increases from n~0 (the log) to n=1, one becomes more sensitive to large earthquakes but loses dynamic range in the deep learning network as measured by mean value/maximum of the input data. We have done a preliminary analysis of $E^{0.25}$ but this paper only presents an analysis of the data using "Log(Energy)" (magnitude).

Note all input properties and predictions (separately for each data category) are independently normalized to have a maximum modulus of 1 over all space and time values. Note that in deep learning values of O(1) are especially significant as that is the value where activation layers introduce nonlinearity into the analysis. The time dependence of different measures of earthquake activity is compared in figure 5 for 2-week time bins. The data is summed over the 2400 spatial bins but averaged as in equations (1) to (5) within each space-time bin. All plots show $m_{bin}$ defined in (2) which we use for most inputs and outputs and we compare with the multiplicity of events with magnitude > 3.29, the multiplicity of all events, $E^{0.25}$ and $E^{0.5}$. Note that $E^{0.5}$ is the "Benioff Strain" [39]–[42], a measure of the accumulating stress or strain.

The $m_{bin}$ curve on each figure is renormalized to have a maximum that is precisely one half that of other plotted quantities. Unsurprisingly $m_{bin}$ has the least drastic time structure and the multiplicity plots are similar in their overall structure to $E^{0.5}$. We largely deal with $m_{bin}$ as its smoother behavior is easier to model. We tried some preliminary analysis with $E^{0.25}$ for targets (defined later) but did not find a good description of data as deep learning doesn't easily describe measurements with a large ratio of maximum to mean. These ratios are recorded in Table 1, where for the model, the individual pixel row is most relevant.

| Ratio Mean/Max for observables | $m_{bin}$ | Full Multiplicity | Multiplicity m > 3.39 | $E^{0.25}$ | $E^{0.5}$ |
|---|---|---|---|---|---|
| **Individual Pixels** | 0.11 | 0.0000046 | 0.000022 | 0.0032 | 0.000016 |
| **Summed over space** | 0.83 | 0.012 | 0.0079 | 0.53 | 0.025 |

**Table 1: Ratios of mean to the maximum of potential observables**

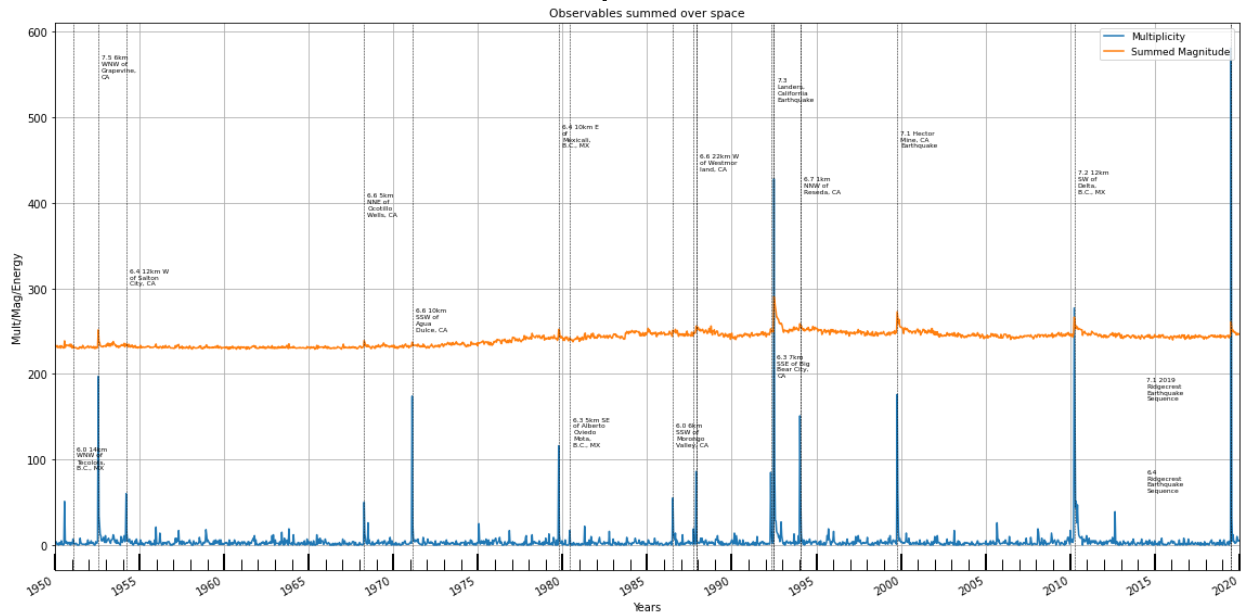

Fig 5a): Time dependence of Multiplicity > 3.29 events compared to $m_{bin}$ of Eq. (2)

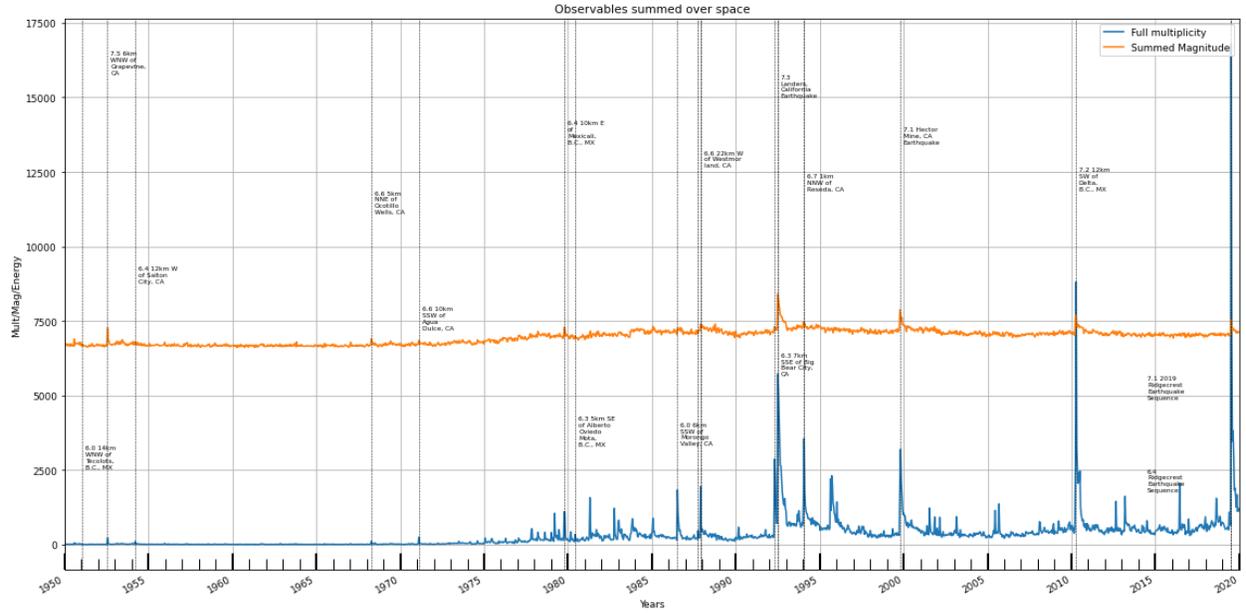

*Fig. 5(b): Time dependence of multiplicity of all events compared to $m_{bin}$ of Eq. (2)*

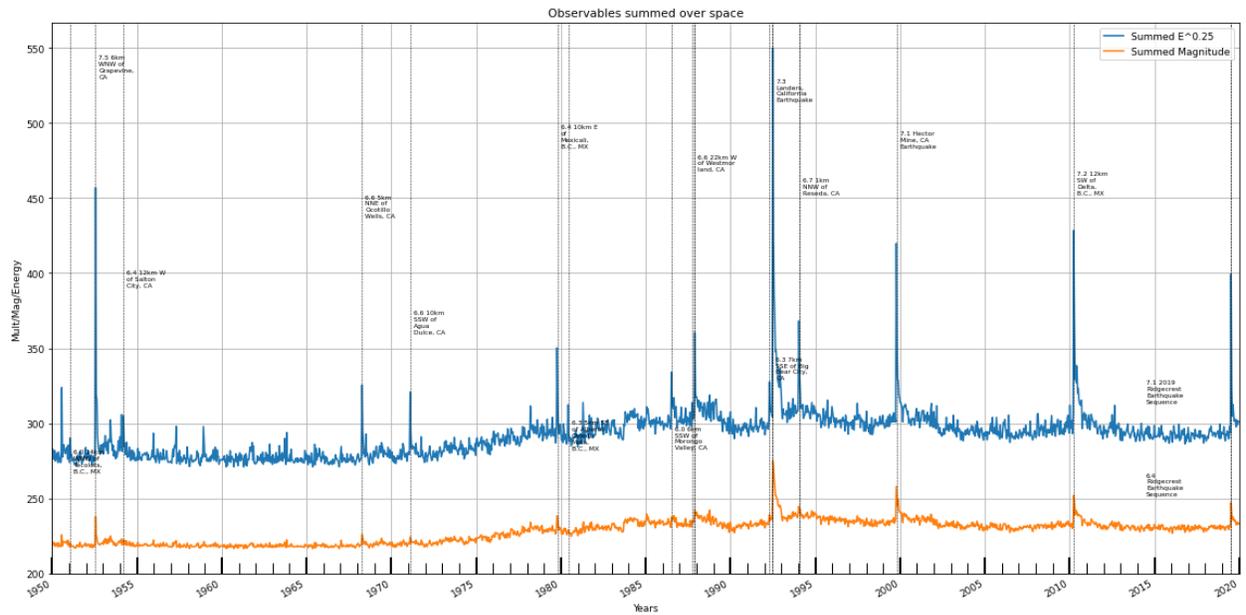

*Fig. 5(c): Time dependence of Energy$^{0.25}$ in fortnightly bins compared to $m_{bin}$ of Eq. (2). Note the suppressed zero in plot.*

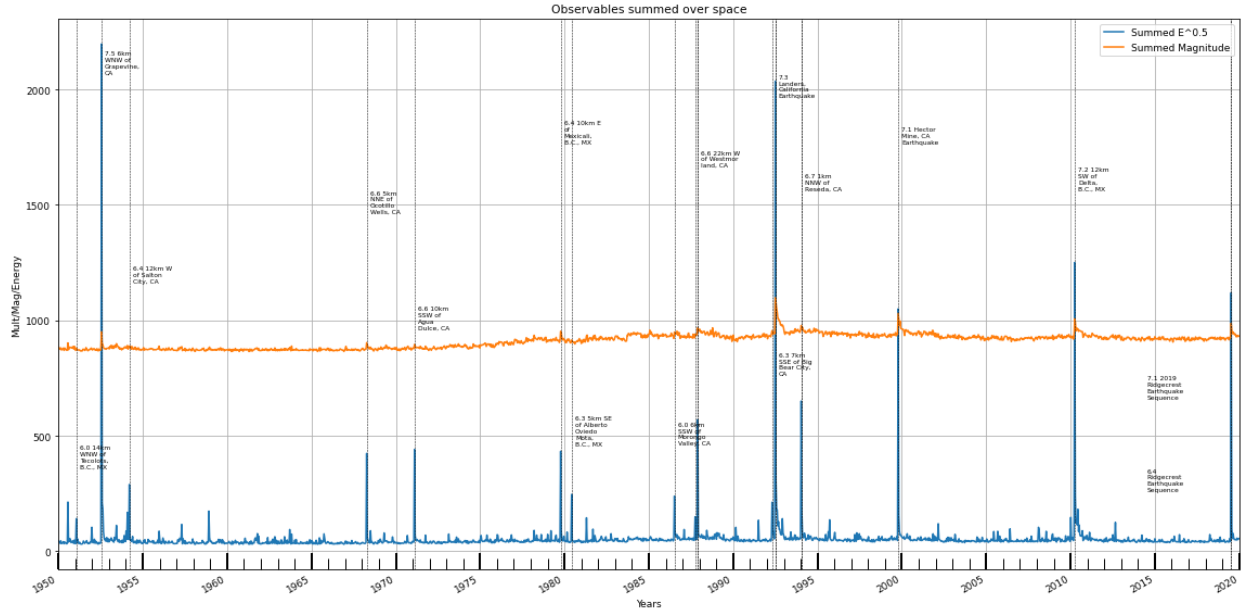

*Fig. 5(d): Time dependence of Energy$^{0.5}$ in fortnightly bins compared to $m_{bin}$ of Eq. (2)*

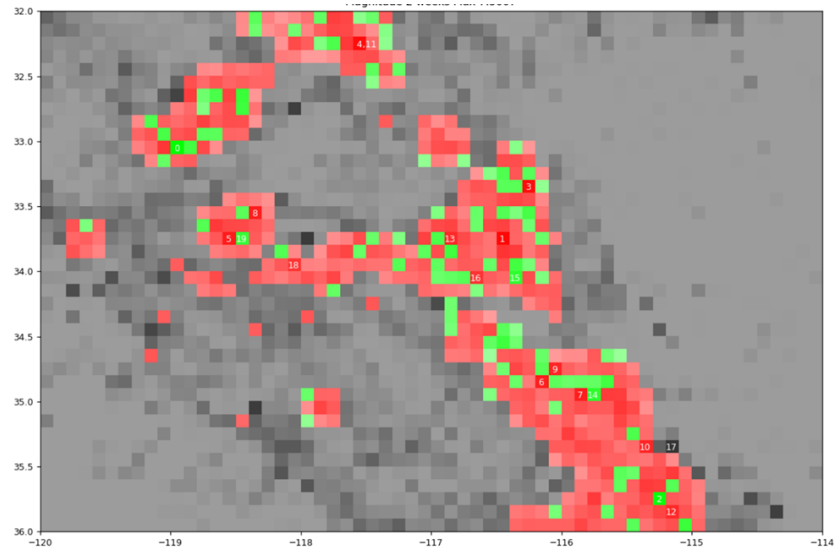

*Fig. 6: The 500 spatial regions (pixels) used in the analysis are divided randomly into 400 red training and 100 green validation/testing pixels with the remaining 1900 regions in shades of grey.*

In the results presented here, we just considered the 500 most active "pixels" among 2400 of the full dataset. These active regions were selected similarly to earlier papers by cuts on the counts of events with magnitude > 3.29. Of these 400 pixels were used for training and 100 for validation and testing. 2400 0.1 by 0.1-degree subregions analyzed (pixels) are illustrated in figure 6 with 400 red as training and 100 green as validation pixels with the pixel intensity corresponding to earthquake activity.

This analysis is built around forward and backward observables using 2-week time units and the $m_{bin}$ observables which are calculated in various time intervals and in the forward and backward directions. $m_{bin}$ (F:Δt,t) is the energy averaged magnitude of equation (2) calculated over the time interval Δt starting at the bin following the time t. $m_{bin}$ (B:Δt,t) is the energy averaged magnitude of equation (2) calculated over the time interval Δt ending at the time t bin. Forward measurements are predicted and backward measurements input. Note that a year prediction $m_{bin}$ (F:Δt=52 weeks,t) can **not** be calculated from the 26 two-week predictions $m_{bin}$ (F:Δt=2 weeks,t to t+25). The energy averaging (roughly a maximum magnitude) does not commute with the averaged prediction. Here earthquake analysis differs from other time series where observables are directly measured event counts.

## 2.2 Earthquake Faults

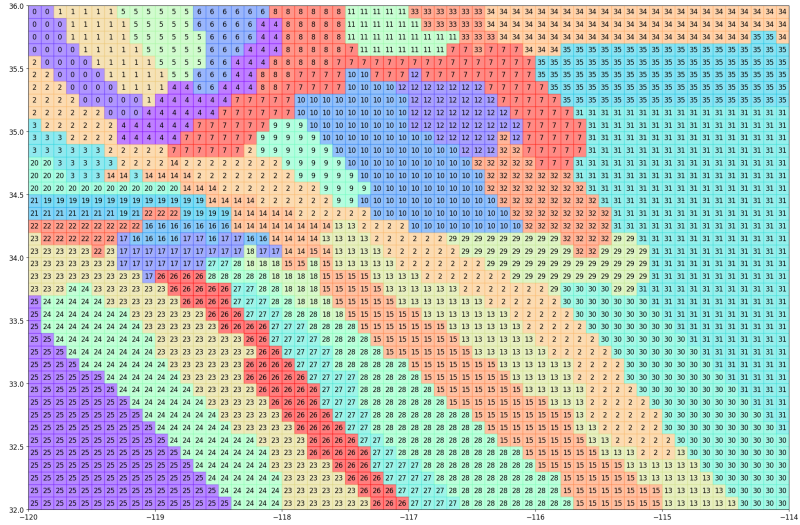

*Fig. 7: Fault Regions with a particular space-filling label for each of the 2400 pixels*

Hundreds of faults have been identified and it is important to understand how to use this information in the models. Visually the faults look like the highways in the related traffic problems but there is not such a clear relation between quakes and faults as a function of time as there are for cars and roads.

The road structure is described by a graph neural net but we didn't use this approach here but rather after much experimentation settled on a simple idea shown in fig. 7. We divided the faults into groups (36 in results presented here) and labeled each region by the index on a space-filling curve running through our region. Then each of the 2400 pixels is given a static label from 0..35 which is used as described below. There are many ways of generating space-filling curves and we chose 4 independent ones and so each pixel ends with four static labels.

## 3 Formulation of Deep Learning for Time Series
### 3.1 Spatial Bag Problem

We introduced the concept of a spatial bag for time series earlier and illustrate this in fig. 8, where we have a set of time series where the spatial distances (e.g. locality) between points is not important; rather they are differentiated by values of properties which can either be static (such as percentages of the population with high blood pressure for medical examples or minimum annual temperature for hydrology catchment studies) or dynamic (such as a local social distancing measure). In other papers, we will discuss problems that combine the features of spatial bags and distance locality where convolutional networks (especially convLSTM) are clearly useful. We tried convLSTM for the Earthquake case but did not find it as effective as the methods reported here.

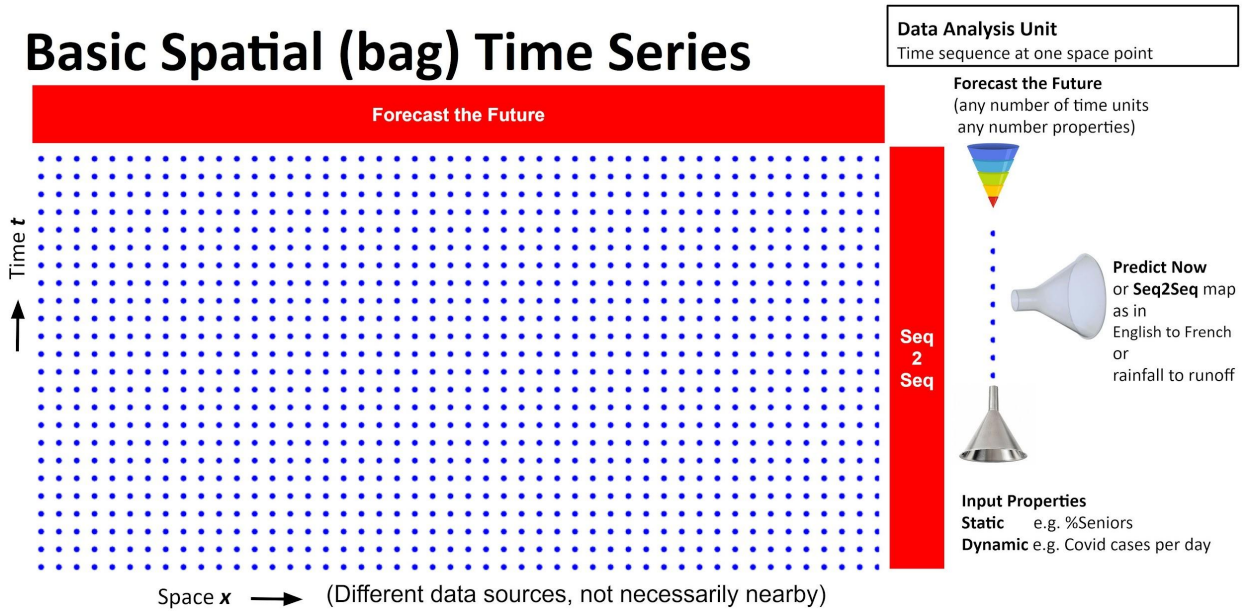

*Fig 8: Illustration of a spatial bag with associated time sequences. t labels time and x location, The Seq2Seq and forecast modes are illustrated and the structure of input data and predictions.*

### 3.2 Deep Learning implementations

Four distinct deep learning implementations will be considered
1) **LSTM** Pure recurrent Neural Network -- two-layer LSTM [43]
2) **TFT** A version of the Google Temporal Fusion Transformer TFT [44]–[46] that uses two distinct LSTM's -- one each for Encoder and Decoder with a temporal (attention-based) transformer
3) **Science Transformer** A hybrid model that was built at the University of Virginia with a space-time transformer for the Decoder and a two-layer LSTM for the encoder.
4) **AE-TCN Joint Model** This is an architecture that combines two models. One is an AutoEncoder (AE) that encodes and decodes the image-like input and output. The other is a Temporal Convolutional Network (TCN) which takes the differences from the input and output of the AutoEncoder and predicts the one-step future loss which is used to nowcast immediate earthquakes in [49, 50].

All of these are programmed using the standard Tensorflow 2 class model. This required some adjustment for TFT where the software is presented [47] in Tensorflow 1 compatibility mode running under Tensorflow 2. NVIDIA has a modern version of the TFT with PyTorch [48]. We modified the TFT in a few ways to allow multiple targets and multi-layer LSTM's. We also use Mean Square Error MSE and not quantiles (mean absolute error) as the loss function. This loss is used identically in all 3 models. MSE appears more appropriate than MAE in this problem as it emphasizes the interesting large earthquake region; MAE weights small earthquakes more than MSE.

### 3.3 The Deep Learning Models in Detail

Deep Learning models are typically constructed as a set of connected layers that calculate forward and back-propagation supported by the well-known PyTorch and Tensorflow frameworks. Each layer has parameters one can change and the user can also vary the order and nature of layers. We have three implementations with rather different structures and sizes. The smallest model is the LSTM with 6 layers shown in figure 9 and 66,594 trainable parameters. Layers used include basic dense, dropout, activation, softmax, add, multiply as well as complex layers such as LSTM. The broad classes we use are dense embedding layer for initial and projection layer for final processing, transformers to identify patterns, and LSTM recurrent networks for time-series processing. In other work, we have used autoencoders and temporal convolutional networks to identify earthquakes as extreme events [49], [50]. We illustrate the simplest network in fig. 9 where grey represents layers and yellow-green data arrays with dimensions. The high-level architectures are compared in fig. 10 while the detailed TFT architecture is well described in [44], [45] and the other two models can be found in [51], [52].

| Network | # Layers | # Trainable parameters | Window Size | Internal Label and online LInk |
|---|---|---|---|---|
| **LSTM** | 6 | 66590 | 13 | EARTHQB-LSTMFullProps2 [53] |
| **TFT** | 73 | 8005334 | 26 | EARTHQ-newTFTv29 [54] |
| **Science Transformer** | 14 | 2339328 | 26 | Y8-EARTHQDGX-Transformer1[55] |

**Table 2: Some Features of the 3 models. The #layers are only for the highest level and nearly all of them are composite and so the full count is much higher.**

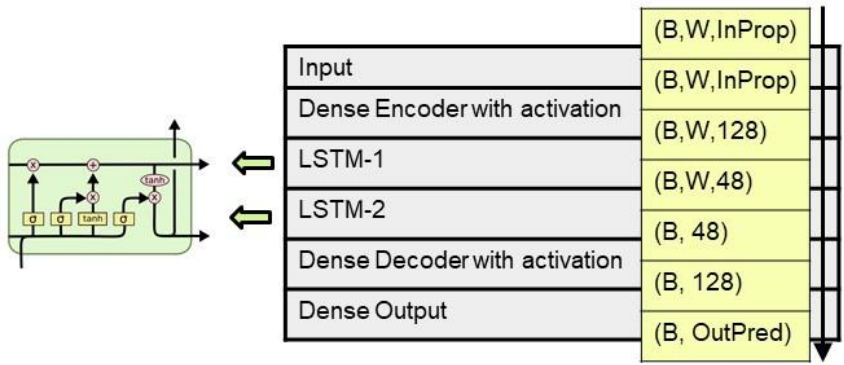

*Fig. 9: The LSTM deep learning network used in this analysis. The 2 LSTM layers are sophisticated compound networks illustrated on the left. B is Batch size (400), W is window size (13); InProp the number of input time series (27) and OutPred the number of predicted time series (24).*

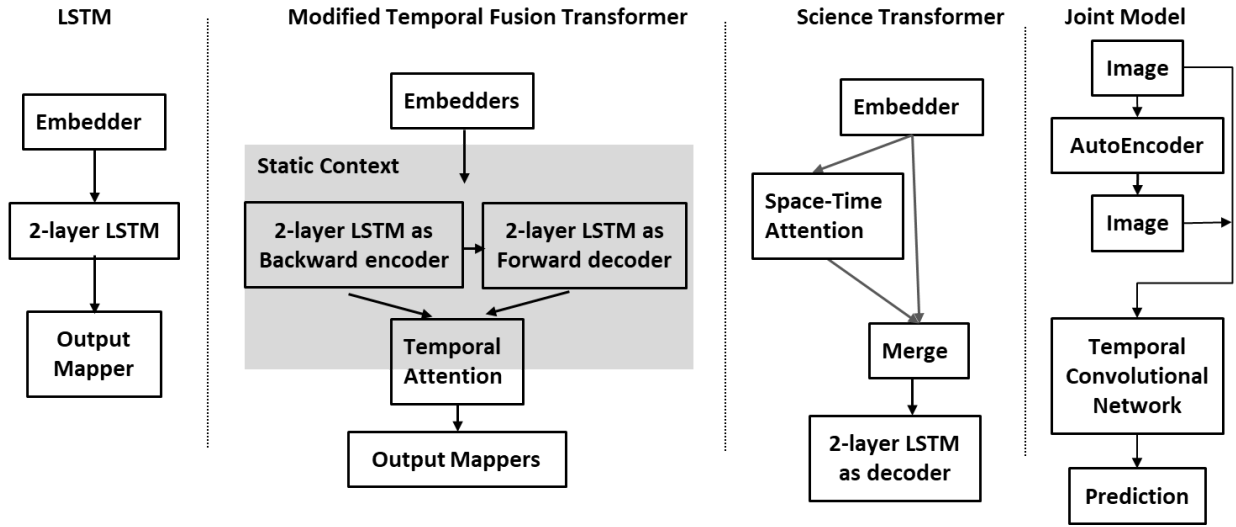

*Fig. 10: High-level Architectures of four deep learning networks*

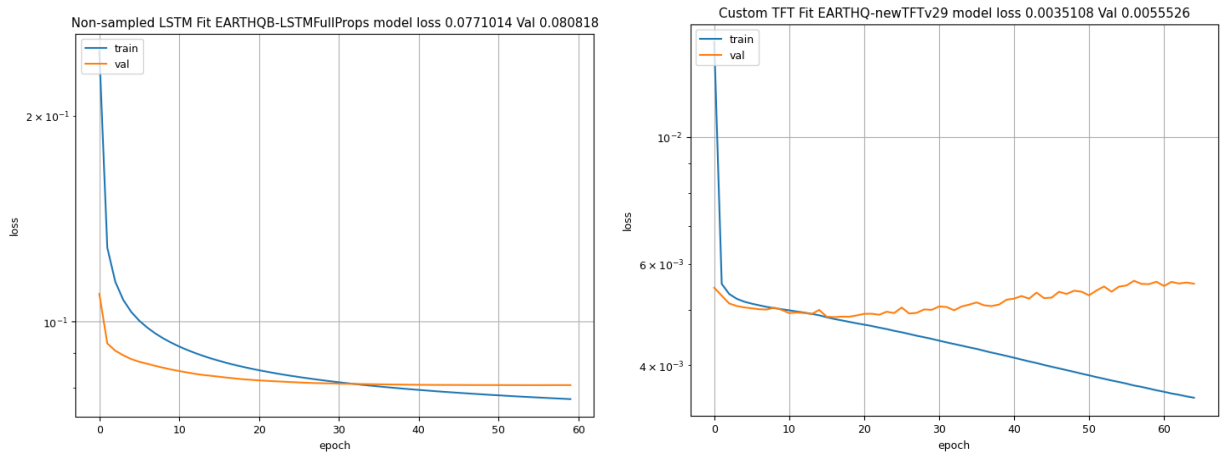

*Fig. 11 Epoch dependence of training and validation loss for LSTM (left) and TFT (right).*

Fig.11 gives typical plots of convergence of two of the models as a function of epoch. The models have different objective functions so absolute values cannot be compared. One can use results at any epoch trading off the training loss (which continues to decrease) with further epochs versus the validation loss which tends to stabilize.

The Temporal Fusion Transformer has two important differences from the other two models. Firstly it uses static variables to provide context (initial values) for several of the time-dependent steps. Secondly, it uses separate input embedders and output mappers for each variable whereas the other two methods use shared embedders and mappers. In future work, these TFT strategies should be considered in the other methods.

### 3.5 The Deep Learning Inputs and Outputs

Deep Learning involves somewhat different inputs and outputs (predictions or targets) to more traditional approaches. Following [45], we identify three major classes of features which are functions of time (time series) for each member of the spatial bag (i.e each pixel in fig. 6) or static characteristics of the bag member. These classes are **Known Inputs**, **Observed Inputs,** and **Targets**. We describe these in general and then specify how they were chosen for each model.

**Observed Inputs** are the basic measured quantities such as magnitude and depth of earthquakes. In the Covid example illustrated in fig. 3, they are daily infections and fatalities but also time-dependent auxiliary variables such as vaccination rate, measures of social distancing, hospitalization rates etc.. In the earthquake case, bin multiplicity counts are auxiliary variables. More formally we divide observed inputs into "(a priori) **unknown inputs**" and **targets** as one typically has variables that are both observed inputs at times up to current time and these need to be predicted and so fall in the target category as well.

**Known Inputs** are an interesting concept that includes both **static features** and time series (**known time-dependent features**). These are parameters that are known in both the past and future whereas observed inputs are only known in the past and need to be nowcast into the future. In Covid and Earthquake examples the only measured known inputs are the static features. In commercial applications, daily signatures of holidays or weekends are interesting inputs. Note these are categorical variables and generally all parameters can be categorical or come from a numerical scale. The latter tend to dominate scientific time series. In our analysis, we made extensive use of **Mathematical Expansions** which are functions of time that it appears natural to express the time dependence in terms of. In general, this is the place where simple models or theoretical intuition can be fed in but there is so far not much experience in exploiting this. As well as feeding in theoretical intuition in "known inputs" one can also feed in sophisticated functions of the observed data -- such of those in our earlier work -- as "observed inputs" allowing the deep learning to identify precursor patterns of future earthquakes.

Fig. 12 shows the modified TFT applied to describe daily Covid data up to November 29, 2021 (extended from [43]). The weekly mathematical "known input" helps the model obtain a good description. The annual periodic variation was used to advantage for describing hydrology data in [52]. We also used Legendre polynomials as known inputs $P_l(cos\theta_{Full})$ where $cos\theta_{Full}(t)$ varies from -1 to +1 over the full-time range of the problem. The results here use the Legendre polynomials from l=0 to 4. Note $cos\theta_{period}(t)$, $sin\theta_{period}(t)$, $P_l(cos\theta_{Full}(t))$ lie between -1 and +1 and fit the deep learning framework constraints quite well as they do not create poor ratios of mean/max of sizes. These are known inputs as they can be calculated for any time value -- past or future.

**Targets** have already been introduced as these are the functions of time that we are trying to predict. They often include all or a subset of the observed inputs as for training they need to be known for times previous to that where the nowcasting was made. However just as we had unknown inputs that were time series which were not predicted, we also sometimes used **Synthetic Targets** that were predicted but not part of input set. These need to be known for "all times" so they can be used in training.

However, whereas inputs are best if they have no missing data, that is not required for targets. These can be missing as we are using an additive loss function (MSE or MAE) and missing targets are just dropped from the sum in the loss function. We use this for predicting many years in the future which are dropped from training at times such that the future measurements are not available. If targets predict $T_{forward}$ time units into the future we choose the maximum training time so the targets with the smallest forward are known but not the others. In the simplest models, targets are the values of observables at the final time. However, we also use synthetic futures when observables are $m_{bin}$ (F:Δt,t) integrals over time periods Δt up to 4 years.

We now summarize the choices that we made for the three networks discussed in this paper. Note for the LSTM and Science Transformer static variables are implemented as "known input time series" with values independent of time. For the TFT, static variables are cleverly used to provide initial context for appropriate layers processing the true time series.

| Static Known Inputs (5) | 4 space-filling curve labels of fault grouping, linear label of pixel |
|---|---|
| Targets (24) | $m_{bin}$ (F:Δt,t) for Δt = 2, 4, 8, 14, 26, 52, 104, 208 weeks. Also for skip 52 weeks and predict next 52; skip 104 and predict next 104. With relative weight 0.25, all the Known inputs and linear label of pixel |
| Dynamic Known Inputs (13) | $P_l(cos\theta_{Full})$ for l=0 to 4<br>$cos\theta_{period}(t), sin\theta_{period}(t)$ for *period = 8, 16, 32, 64* |
| Dynamic Unknown Inputs (9) | Energy-averaged Depth, Multiplicity, Multiplicity m>3.29 events $m_{bin}$ (B:Δt,t) for Δt = 2, 4, 8, 14, 26, 52 weeks |

**Table 3: Input and Output variables for the LSTM and Science Transformer**

| Static Known Inputs (5) | 4 space-filling curve labels of fault grouping, linear label of pixel |
|---|---|
| Targets (4) | $m_{bin}$ (F:Δt,t) for Δt = 2, 14, 26, 52 weeks. Calculated for t-52 to t for encoder and t to t+52 weeks for decoder in 2 week intervals. 104 predictions per sequence. |
| Dynamic Known Inputs (13) | $P_l(cos\theta_{Full})$ for l=0 to 4<br>$cos\theta_{period}(t), sin\theta_{period}(t)$ for *period = 8, 16, 32, 64* |
| Dynamic Unknown Inputs (9) | Energy-averaged Depth, Multiplicity, Multiplicity m>3.29 events $m_{bin}$ (B:Δt,t) for Δt = 2, 4, 8, 14, 26, 52 weeks |

**Table 4: Input and Output variables for the modified Temporal Fusion Transformer**

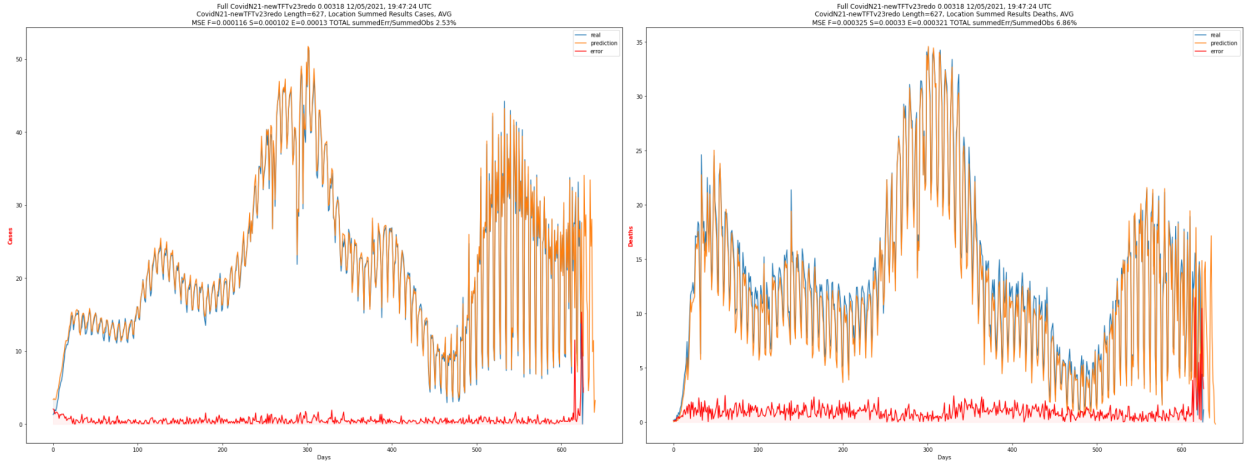

Fig 12: Modified TFT description of 500 most populous counties in the USA for Covid daily infection and fatality data showing strong weekly periodic structure from early 2020 to November 2021.

The deep learning models are perhaps the important glamorous parts of the analysis but substantial data engineering is needed to pre-process the input data and visualize the results. This hard work is described in [52] and can be viewed in the online Jupyter notebooks.

## 4 Transformer and Attention Technologies

### 4.1 Using Transformers for Spatial Bags

The deep learning methods use approaches that were largely originally developed for natural language processing.

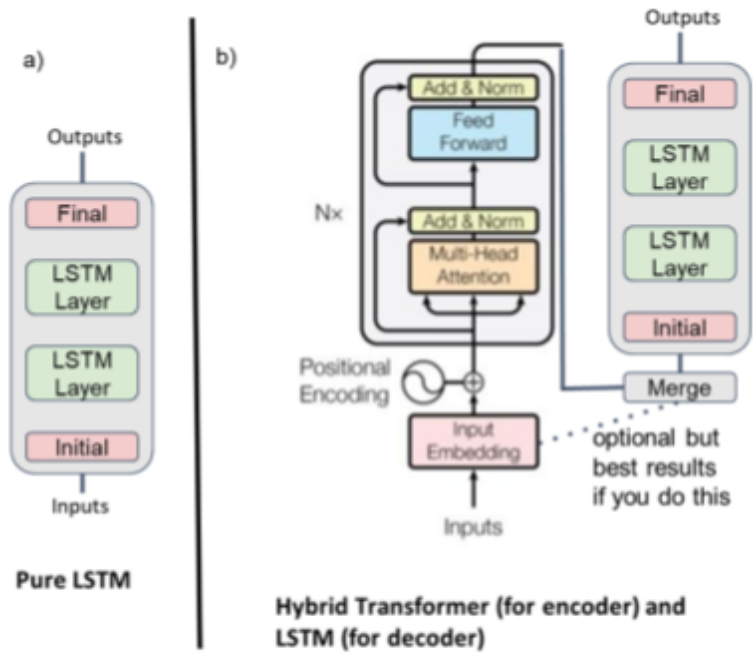

Fig. 13: a) Pure LSTM and b) Hybrid Transformer model discussed in this paper. The latter has an attention-based decoder and an LSTM-based final encoder

Recurrent Neural Networks RNN explicitly focus on sequences and pass them through a common network with pretty subtle features. RNNs are designed to gather a history which allows the time dependence to be remembered Attention-based methods [56] are more modern and perhaps somewhat easier to understand as attention is a simple idea.

NLP is basically a classification problem (look up tokens in a context-sensitive dictionary) whereas (science) tends to be numerical and so it is not immediately obvious how to use attention in technical time series and we describe one possible approach in this paper. There have been a few studies of transformer architectures for numerical time series such as [57]–[63] but there is not a large literature.

Attention [64], [65] means that you "learn" structure from other related data and look for patterns using a simple "dot-product" mechanism discussed later matching structure of different sequences; there are other approaches to match patterns which is a good topic for future work. Here we use a simple attention mechanism in an initial decoder but use a recurrent net LSTM for the encoder as shown in fig. 5b). Such mixtures have been investigated and compared [66], [67]. We compare the two architectures shown in a) and b) of fig. 13; the pure LSTM used in sec. 2 and the science transformer

### 4.2 Scaled Dot-Product Attention and the Vectors Q K V

The basic item for LSTM and Transformer is the same; a space point with a time sequence with each time in the sequence having a set of static and dynamic values. In an LSTM the sequence is "hidden" and you have to unroll the recurrent network to see it. However in the transformer, the different time values in a sequence are treated directly and so each item contains W terms (W is the size of time sequence), Each term is embedded in an input layer and then mapped by 3 different layers into vectors Q (query) K (key) V (value).

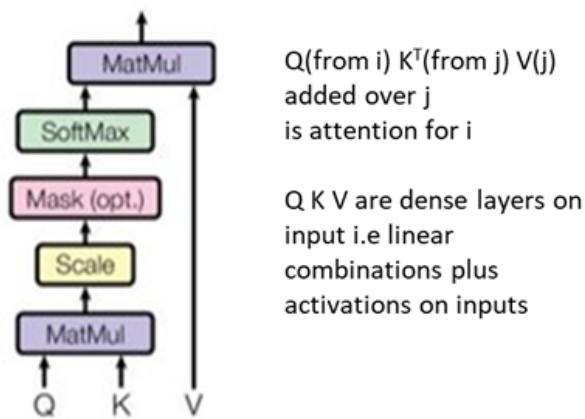

Fig. 14: The Q K V layers of the Attention Mechanism.

As shown in fig. 14, one matches terms i and j by calculating $Q(i)K^T(j)$ and ranking with a softmax step. This multiplies the characteristic vector V(j) of this pattern and the total attention A(i) for item i, is calculated as a weighted sum over the values V(j). There are several different attention "heads" (networks generating Q K V) in each step and the whole process is repeated in multiple encoder layers. The result of the encoder step is considered separately for each item (each time in a time sequence at a given location) and the embedded input of this layer is combined with the attention as input to the LSTM decode step.

### 4.3 Choosing the group of items over which Attention Calculated

In natural language processing, you look for patterns among neighboring sentences but for science time series you can have larger regions as spatial bags have no locality. This leads to many choices as to the space over which attention is calculated as one can't realistically consider all items simultaneously.

Suppose we have $N_{loc}$ locations; each with $N_{seq}$ sequences of length W. Then the space to be searched has size $N_{loc} \cdot N_{seq} \cdot W$ which is too large. In COVID-19 example of fig. 12 [43] $N_{loc}$= 500, $N_{seq}$ ~ 700 and W up to 13. In the hydrology example in [52], $N_{loc}$= 671, $N_{seq}$ ~ 7000 and W up to 270. The next subsection describes the 3 search strategies we have looked at: and depicted in fig. 15. There is

- Temporal search: points in sequence for fixed location
- Spatial search: locations for a fixed position in the sequence
- Full search: complete location-sequence space

One will need to sample items randomly as only a small fraction of space is looked at in one attention step whatever method is used. Note that in all cases we used a batch size of 1 as the attention space was effectively the batch. Actually, in the LSTM stage, the different locations in the attention space were considered separately and the attention search space became the batch. In the work reported here the attention space (and batch size) was set to $N_{loc}$ but this is not required and would not work in some cases with large values of $N_{loc}$.

Even in the examples considered here, the search space can get so large that one needs to address memory size issues. Note the encoding step of the transformer is many matrix multiplications and gets excellent GPU performance and typically LSTM decoder would be a significant part of the compute time and so the addition of attention is not a major performance hurdle although it does require a large GPU memory. The TFT only uses temporal attention separately for each space point and so only the science transformer needs major (80 GB of A100) GPU memory.

a) Temporal: At each location look over time window $O(NW^2)$

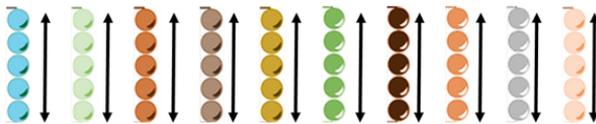

b) Spatial: Look across items at fixed position in time window $O(N^2W)$

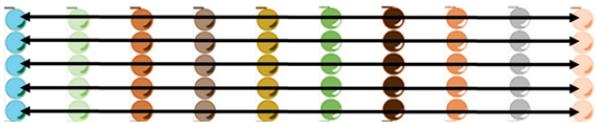

c) Full: Look over all space time windows in batch $O(N^2W^2)$

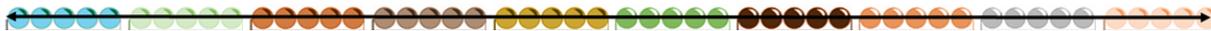

Fig. 15: Three search strategies discussed in paper. $N=N_A$ is the number of items considered in attention search- each labeled by a location and starting time - and each a window of length W. W=5 in diagram. $N_A = N_B = N_{loc}$ here.

An epoch contains $N_{loc} \cdot N_{seq}$ sequences each of length W and consisting of static and dynamic variables characteristic of location. For an LSTM, you would have a batch that consists of $N_B \sim N_{loc}$ sequences. The number of batches in an epoch is approximately $N_{seq}$. For a transformer, the batch is currently unwrapped as discussed above and used as the attention space. Memory usage or compute issues could require a different strategy separately considering batch and attention space.

Note the attention space choice implies different initial shuffling to form batches and also different prediction stage approaches. For spatial and temporal searches one can keep all locations at a particular time value together in both forming batches and in calculating prediction. For the full search the complete set of $N_{loc}.N_{seq}$ sequences must be shuffled. In practice, we combined the space-time and pure time searches and accumulated the results of these two attention searches.

## 5 Initial Earthquake Nowcasting results

These models give very detailed predictions over time and space and we can only illustrate them in this paper. We give a global summary with the Nash Sutcliffe Efficiency and a sample of time-dependent results in figures of earthquake activity $m_{bin}$ (F:Δt,t) .
The Nash Sutcliffe Efficiency NSE is given in equation 1 where we use the normalized NSE, NNSE = 1/(2-NSE) which runs between 0 (bad) and 1 (perfection). Table 5 gives the results are for $m_{bin}$ summed over all 400/100 locations. Note all time interval bins use the forward $m_{bin}$ described in section 2 and equation (2)  Figures 16-19 show the nowcasts for training and validation sets for a selection of time intervals: 2 weeks, one year, two years and four years. In this initial study, the 3 methods give qualitatively similar results which have strong promising potential for useful nowcasts.

The time period listed corresponds to the forward prediction of that length. The TFT only uses backward prediction but does this for time steps reaching 26 weeks into the future so the backward one-year value at t+26 fortnights is the forward prediction at time t. It is not clear if the detailed 2-week predictions of TFT into the future (not given in other models) are valuable. We noted before that one cannot use these predictions to calculate cumulative observables as energy-averaging and finding mean are not commutative. The current TFT implementation was not set up to nowcast all the time intervals covered by the other models and so this model is absent in some cells of Table 5 and in the later figures for 2-year and 4-year nowcasts. This is not intrinsic to the TFT model and will be addressed in future work using larger memory machines to train networks.

|  | Normalized Nash Sutcliffe Efficiency NNSE | | | | | |
|---|---|---|---|---|---|---|
| Time Period | LSTM Train | TFT Train | Science Transformer Train | LSTM Validation | TFT Validation | Science Transformer Validation |
| **2 weeks** | 0.903 | 0.915 | 0.893 | 0.868 | 0.875 | 0.856 |
| **4 weeks** | 0.895 |  | 0.916 | 0.867 |  | 0.884 |
| **8 weeks** | 0.886 |  | 0.913 | 0.866 |  | 0.881 |
| **14 weeks** | 0.924 | 0.963 | 0.919 | 0.893 | 0.905 | 0.881 |
| **26 weeks** | 0.946 | 0.956 | 0.954 | 0.897 | 0.892 | 0.896 |
| **52 weeks** | 0.919 | 0.957 | 0.955 | 0.861 | 0.871 | 0.876 |
| **104 weeks** | 0.923 | | 0.937 | 0.853 | | 0.83 |
| **208 weeks** | 0.935 | | 0.921 | 0.811 | | 0.77 |

**Table 5. Summary of Nowcasting results in terms of predicted Nash Sutcliffe Efficiencies**

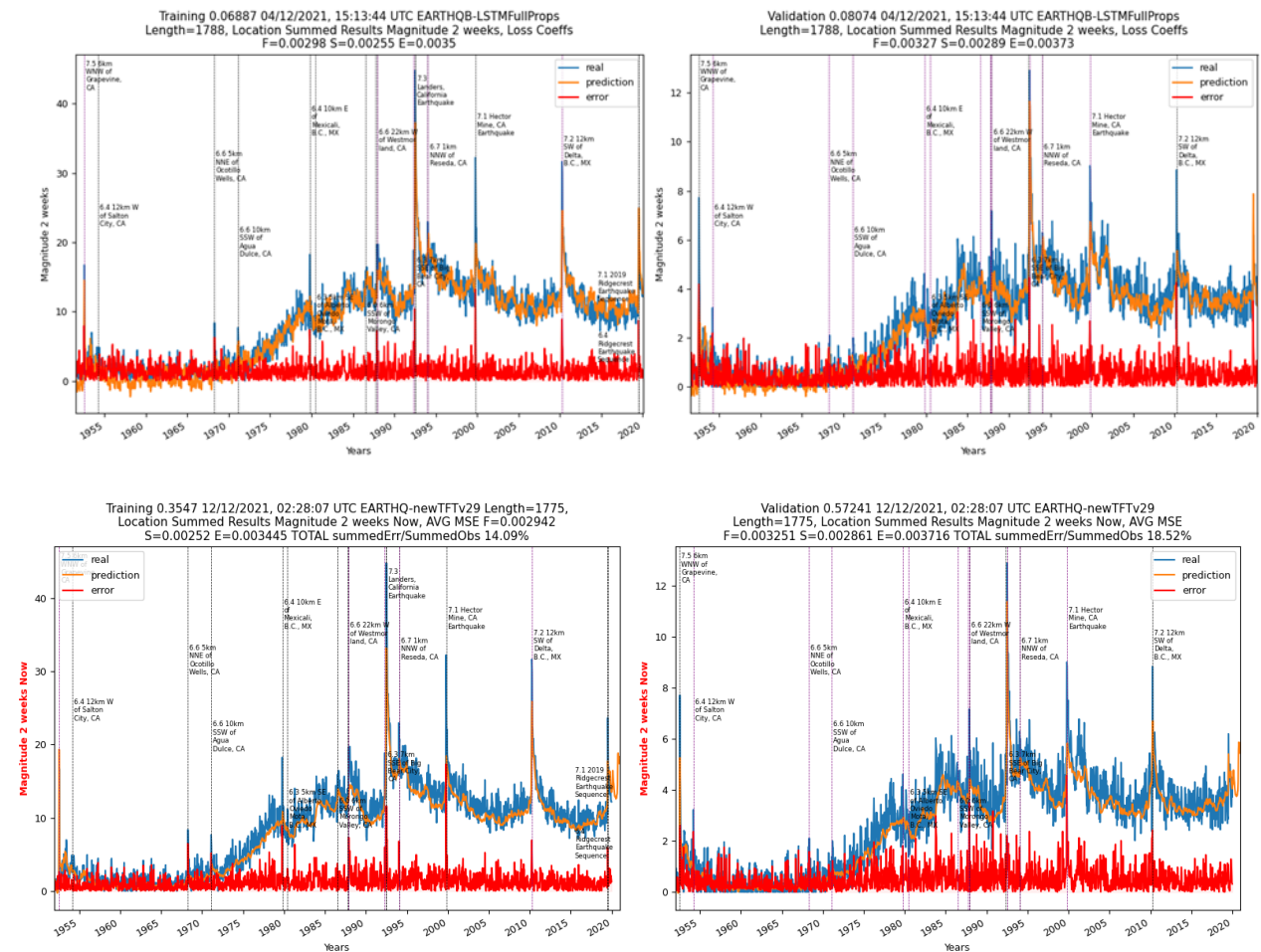

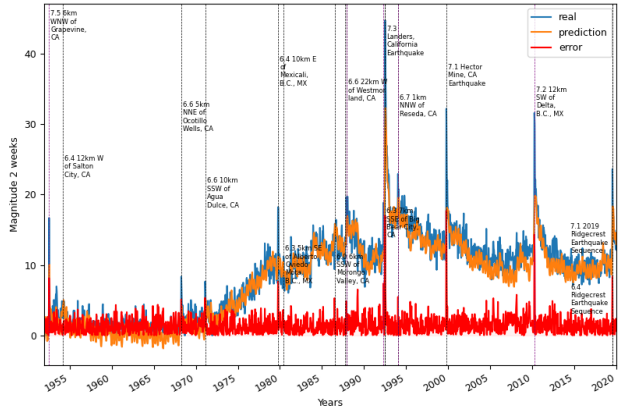
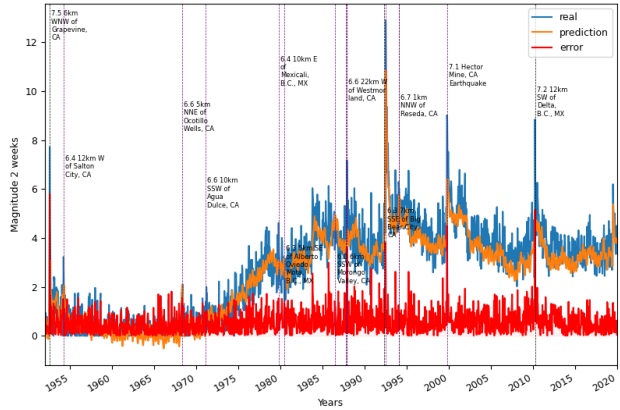

*Fig 16 $m_{bin}$ (F:Δt, t) results for Δt=2 weeks (training on left, validation on the right) are given for 3 methods in the order: LSTM, TFT, Science Transformer.*

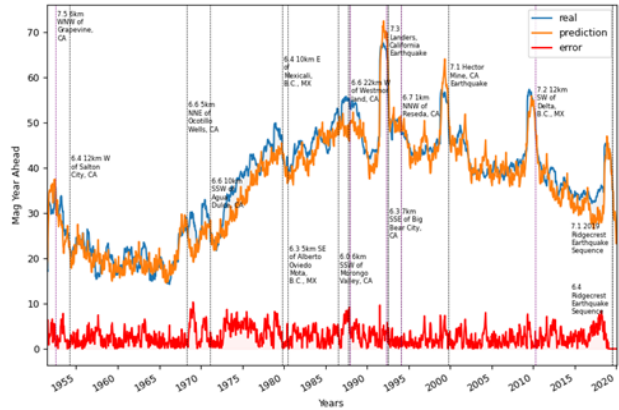
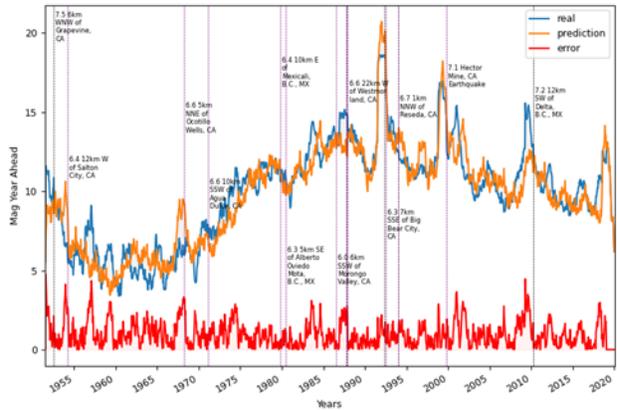
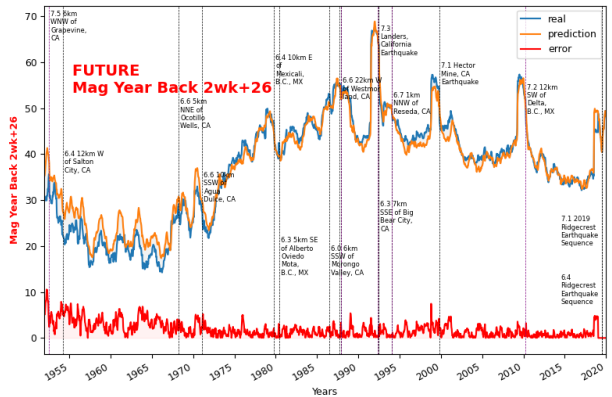
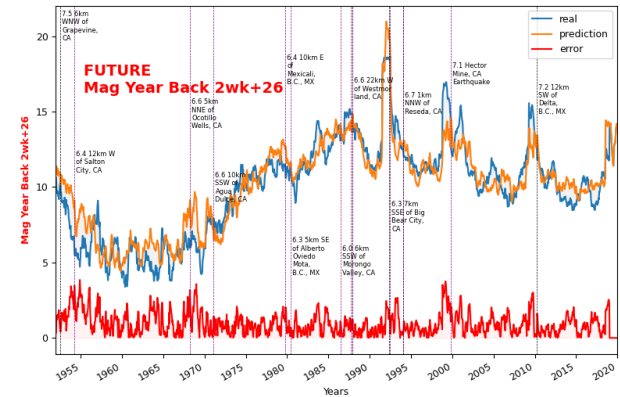

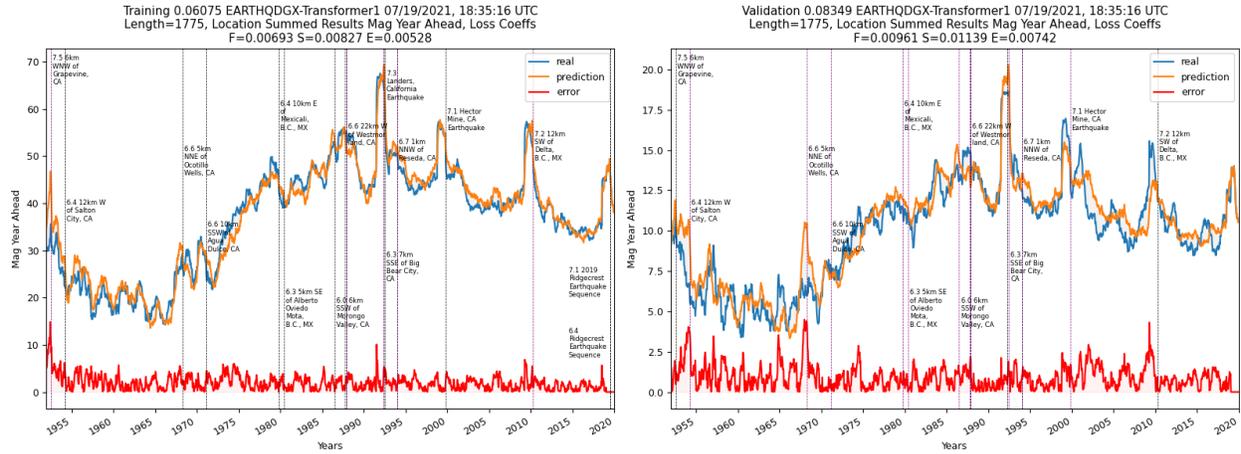

*Fig 17 $m_{bin}$ (F:Δt, t) results for Δt=52 weeks (training on left, validation on the right) are given for 3 methods in the order: LSTM, TFT, Science Transformer. Note TFT results are formed using $m_{bin}$ (F:Δt, t) = $m_{bin}$ (B:Δt, t+Δt)*

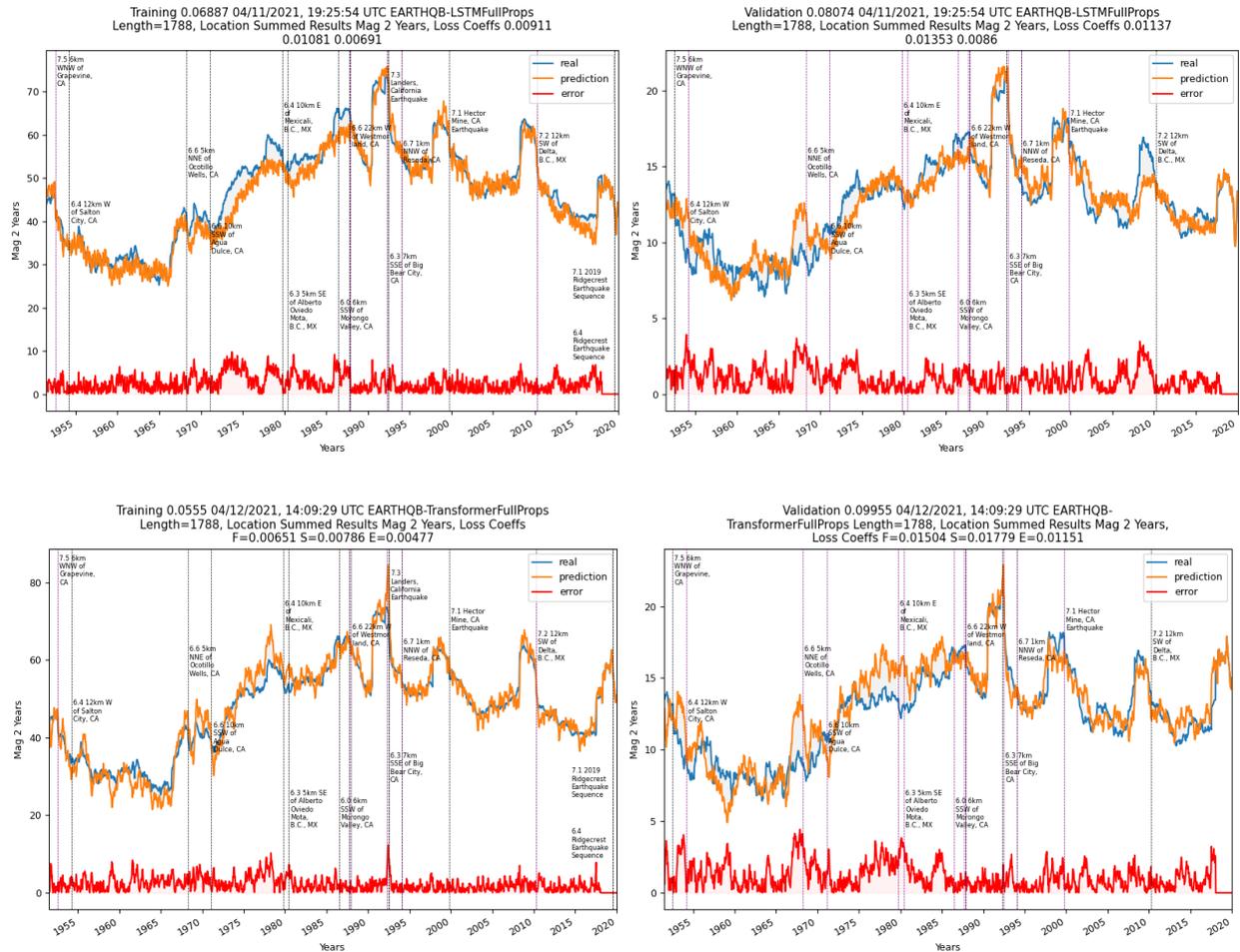

*Fig. 18 $m_{bin}$ (F:Δt, t) results for Δt=104 weeks (training on left, validation on the right) are given for 2 methods in the order: LSTM, Science Transformer.*

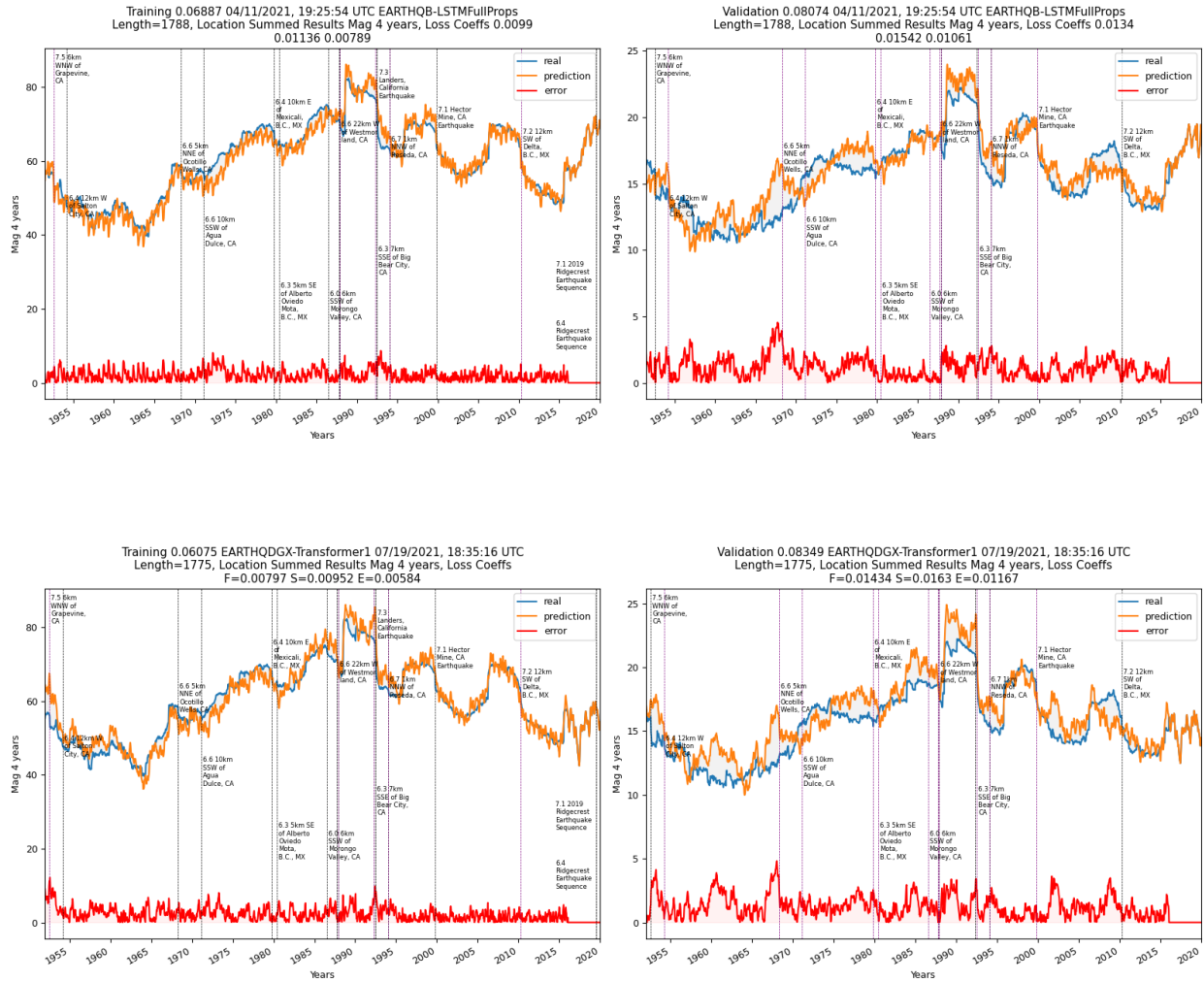

*Fig. 19 $m_{bin}$ (F:Δt, t) results for Δt=208 weeks (training on left, validation on the right) are given for 2 methods in the order: LSTM, Science Transformer.*

## 5.2 Nowcasting Extreme Events with AE-TCN Joint Model

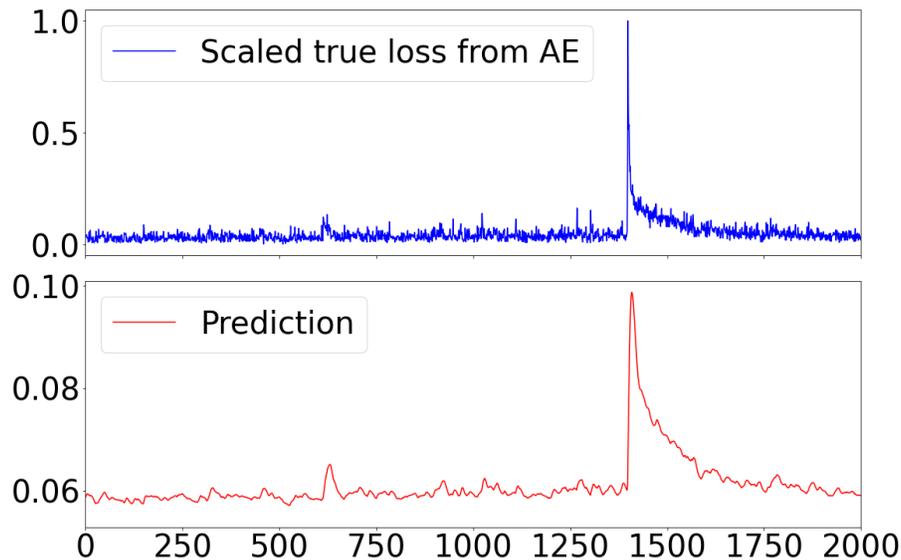

*Fig. 20 An empirical study with the AE (autoencoder)-TCN (Temporal convolutional network) joint model.*

The AE-TCN joint model is built as described in previous research [49], [50] and an overview architecture is shown earlier in Figure 10. This research is an empirical study for predicting extreme shocks as major events controlled by a pre-configured threshold value. Fig 20 illustrates a result from a test set covering around 2,000 data points, where each data point is an image-like snapshot of Southern California with pixels representing shock energy. We calculate the loss of the AutoEncoder as the mean absolute error from the reconstruction to the input. In Figure 5, the top sub-figure is the min-max scaled loss using the test dataset. The bottom sub-figure of Figure 20 is the model prediction using the test dataset. A high value in prediction indicates the corresponding event is large. In the joint modeling, an autoencoder is responsible for encoding and decoding image input and a temporal convolutional network takes the differences from the autoencoder and predicts high loss, which indicates large events one step ahead.

## 6 Conclusions

Above we have presented an initial analysis of the use of deep learning for nowcasting earthquakes. We believe the methods are promising and have broad applicability across earth science which offers many examples of geospatial time series. We have identified a class -- spatial bags - where different space points are linked by different values of static labels and not directly by the geometry, We see very complex networks with from over 60,000 to 8 million trainable (unknown) parameters. We use the data to develop the model which then replaces the physics description with these hidden variables.

Note a feature of the gradient descent optimizers is that you can use a redundant parametrization and still converge to a reasonable description. You will over-train with lots and lots of parameters but by comparing training and validation loss find a good description presented here. This paper is not a definitive study, as it is probing somewhat uncharted territory. In particular, we have made very many heuristic choices that need further investigation -- or one could say there are many hyperparameters in this problem that need to be explored.

Areas to explore include time-unit (2 weeks here), the choice of known inputs, the many parameters defining the networks, the targets, and the basic variable from $m_{bin}$ to Energy E and its powers. We also should explore the choice of the validation set; should it be gotten from dividing in space (as here) or time? We earlier noted that the TFT had several clever ideas that could be used in the LSTM and Science Transformer examples. Further for the transformer should one use temporal patterns as in TFT and AE-TCN or the space-time choice made in the Science Transformer.

We intend to build a benchmark set of time series datasets and reference implementations as playing the same role for time series that ImageNet, ILSVRC, and AlexNet played for images. The different implementations establish best practices and can be used in different applications as identified already [68] in finance, networking, security, monitoring of complex systems from Tokamaks[69] to operating systems, and environmental science. This work will be performed in the MLCommons Science Data working group [70]. We intend to combine the open datasets and clean reference implementations available in MLCommons [71] with documentation and tutorials which will allow MLCommons benchmarks to encourage the broad community to study these examples and use the ideas in other applications as well as improving on our base reference implementations.

**Acknowledgments**
The work of GCF and BF is partially supported by the National Science Foundation (NSF) through awards CIF21 DIBBS 1443054, CINES 1835598, and Global Pervasive Computational Epidemiology 1918626. We thank Tony Hey, Jeyan Thiyagalingam, Lijiang Guo, Gregor von Laszewski, Niranda Perera, Saumyadipta Pyne, Judy Fox, Xinyuan Huang, Russell Hofmann, Keerat Singh, JCS Kadupitiya, and Vikram Jadhao for great discussions. Sercan Arik and the Google TFT team were very helpful. We are grateful to Cisco University Research Program Fund grant 2020-220491 for supporting this research. The inspiration of MLCommons has been essential to guide our research. Research by JBR has been supported by a grant from the US Department of Energy to the University of California, Davis, number DoE grant number DE-SC0017324, and by a grant from the National Aeronautics and Space Administration to the University of California, Davis, number NNX17AI32G. Portions of the research by Andrea Donnellan were carried out at the Jet Propulsion Laboratory, California Institute of Technology, under a contract with the National Aeronautics and Space Administration.

**References**

[1] John Rundle and Geoffrey Fox, "Computational Earthquake Science," *Comput. Sci. Eng.*, vol. 14, no. 5, pp. 7–9, 2012 [Online]. Available: http://doi.ieeecomputersociety.org/10.1109/MCSE.2012.94
[2] J. B. Rundle, S. Stein, A. Donnellan, D. L. Turcotte, W. Klein, and C. Saylor, "The complex dynamics of earthquake fault systems: new approaches to forecasting and nowcasting of earthquakes," *Rep. Prog. Phys.*, vol. 84, no. 7, p. 076801, May 2021, doi: 10.1088/1361-6633/abf893. [Online]. Available: https://iopscience.iop.org/article/10.1088/1361-6633/abf893/meta. [Accessed: 09-Dec-2021]
[3] T. Hey and A. Trefethen, "The Fourth Paradigm 10 Years On," *Informatik Spektrum*, vol. 42, no. 6, pp. 441–447, Jan. 2020, doi: 10.1007/s00287-019-01215-9. [Online]. Available: https://doi.org/10.1007/s00287-019-01215-9

[4]  Chaopeng Shen, Penn State University, "D2 2020 AI4ESS Summer School: Recurrent4Neural Networks and LSTMs." [Online]. Available: https://www.youtube.com/watch?v=vz11tUgoDZc. [Accessed: 01-Jul-2020]

[5]  Kratzert, Frederik, "CAMELS Extended Maurer Forcing Data." [Online]. Available: https://www.hydroshare.org/resource/17c896843cf940339c3c3496d0c1c077/. [Accessed: 14-Jul-2020]

[6]  Frederik Kratzert, "Catchment-Aware LSTMs for Regional Rainfall-Runoff Modeling GitHub." [Online]. Available: https://github.com/kratzert/ealstm_regional_modeling. [Accessed: 14-Jul-2020]

[7]  F. Kratzert, D. Klotz, G. Shalev, G. Klambauer, S. Hochreiter, and G. Nearing, "Towards learning universal, regional, and local hydrological behaviors via machine learning applied to large-sample datasets," *Hydrology & Earth System Sciences*, vol. 23, no. 12, 2019 [Online]. Available: https://arxiv.org/abs/1907.08456

[8]  N. Addor, A. J. Newman, N. Mizukami, and M. P. Clark, "The CAMELS data set:Catchment attributes and meteorology for large-sample studies," *Hydrol. Earth Syst. Sci.*, vol. 21, no. 10, p. 21, Oct. 2017, doi: 10.5194/hess-21-5293-2017. [Online]. Available: https://ueaeprints.uea.ac.uk/id/eprint/65434/. [Accessed: 05-Jul-2020]

[9]  M. A. Sit, B. Z. Demiray, Z. Xiang, G. Ewing, Y. Sermet, and I. Demir, "A Comprehensive Review of Deep Learning Applications in Hydrology and Water Resources," 2020 [Online]. Available: https://eartharxiv.org/xs36g/

[10] Yan Liu, "Artificial Intelligence for Smart Transportation Video." [Online]. Available: https://slideslive.com/38917699/artificial-intelligence-for-smart-transportation. [Accessed: 08-Aug-2019]

[11] H. Yao *et al.*, "Deep multi-view spatial-temporal network for taxi demand prediction," in *Thirty-Second AAAI Conference on Artificial Intelligence*, 2018 [Online]. Available: https://www.aaai.org/ocs/index.php/AAAI/AAAI18/paper/viewPaper/16069

[12] X. Geng *et al.*, "Spatiotemporal multi-graph convolution network for ride-hailing demand forecasting," in *2019 AAAI Conference on Artificial Intelligence (AAAI'19)*, 2019 [Online]. Available: http://www-scf.usc.edu/~yaguang/papers/aaai19_multi_graph_convolution.pdf

[13] Y. Wu and H. Tan, "Short-term traffic flow forecasting with spatial-temporal correlation in a hybrid deep learning framework," *arXiv [cs.CV]*, 03-Dec-2016 [Online]. Available: http://arxiv.org/abs/1612.01022

[14] Geoffrey Fox, "Deep Learning for Spatial Time Series," 17 November 2020 [Online]. Available: http://dsc.soic.indiana.edu/publications/Deep%20Learning%20for%20Spatial%20Time%20Series.pdf

[15] Geoffrey Fox, John Rundle, Bo Feng, "Study of Earthquakes with Deep Learning," presented at the Frankfurt Institute for Advanced Study Seismology & Artificial Intelligence Kickoff Workshop, Virtual, 2021 [Online]. Available: https://docs.google.com/presentation/d/1nTM-poaFzrT_KBB1J7BlZdOMEIMTu-s48mcBA5DeP30/edit#slide=id.g7a25695c64_0_0

[16] H. Guan, L. K. Mokadam, X. Shen, S.-H. Lim, and R. Patton, "FLEET: Flexible Efficient Ensemble Training for Heterogeneous Deep Neural Networks," in *Proceedings of Machine Learning and Systems 2020*, 2020, pp. 247–261.

[17] C. H. Scholz, *The Mechanics of Earthquakes and Faulting*. Cambridge University Press, 2019 [Online]. Available: https://play.google.com/store/books/details?id=ynWIDwAAQBAJ

[18] J. B. Rundle and A. Donnellan, "Nowcasting earthquakes in southern California with machine learning: Bursts, swarms, and aftershocks may be related to levels of regional tectonic stress," *Earth Space Sci.*, vol. 7, no. 9, Sep. 2020, doi: 10.1029/2020ea001097. [Online]. Available: https://onlinelibrary.wiley.com/doi/10.1029/2020EA001097

[19] J. B. Rundle and A. Donnellan, "Nowcasting earthquakes in southern California with

machine learning:Bursts, swarms and aftershocks may reveal the regional tectonic stress," *Earth and Space Science Open Archive*, Earth and Space Science Open Archive, 19-Jan-2020 [Online]. Available: http://www.essoar.org/doi/10.1002/essoar.10501945.1

[20] J. B. Rundle, A. Donnellan, G. Fox, and J. P. Crutchfield, "Nowcasting Earthquakes by Visualizing the Earthquake Cycle with Machine Learning: A Comparison of Two Methods," *Surv. Geophys.*, Aug. 2021, doi: 10.1007/s10712-021-09655-3. [Online]. Available: https://doi.org/10.1007/s10712-021-09655-3

[21] J. B. Rundle, D. L. Turcotte, A. Donnellan, L. Grant Ludwig, M. Luginbuhl, and G. Gong, "Nowcasting earthquakes," *Earth Space Sci.*, vol. 3, no. 11, pp. 480–486, Nov. 2016, doi: 10.1002/2016ea000185. [Online]. Available: http://doi.wiley.com/10.1002/2016EA000185

[22] J. B. Rundle, M. Luginbuhl, A. Giguere, and D. L. Turcotte, "Natural Time, Nowcasting and the Physics of Earthquakes: Estimation of Seismic Risk to Global Megacities," in *Earthquakes and Multi-hazards Around the Pacific Rim, Vol. II*, C. A. Williams, Z. Peng, Y. Zhang, E. Fukuyama, T. Goebel, and M. R. Yoder, Eds. Cham: Springer International Publishing, 2019, pp. 123–136 [Online]. Available: https://arxiv.org/pdf/1709.10057

[23] J. B. Rundle, A. Giguere, D. L. Turcotte, J. P. Crutchfield, and A. Donnellan, "Global Seismic Nowcasting With Shannon Information Entropy," *Earth Space Sci*, vol. 6, no. 1, pp. 191–197, Jan. 2019, doi: 10.1029/2018EA000464. [Online]. Available: http://dx.doi.org/10.1029/2018EA000464

[24] J. B. Rundle, W. Klein, K. Tiampo, and S. Gross, "Linear pattern dynamics in nonlinear threshold systems," *Phys. Rev. E*, vol. 61, no. 3, pp. 2418–2431, Mar. 2000, doi: 10.1103/PhysRevE.61.2418. [Online]. Available: https://link.aps.org/doi/10.1103/PhysRevE.61.2418

[25] J. C. Savage, "Principal component analysis of geodetically measured deformation in Long Valley caldera, eastern California, 1983-1987," *J. Geophys. Res.*, vol. 93, no. B11, pp. 13297–13305, Nov. 1988, doi: 10.1029/jb093ib11p13297. [Online]. Available: http://doi.wiley.com/10.1029/JB093iB11p13297

[26] K. F. Tiampo, J. B. Rundle, S. McGinnis, S. J. Gross, and W. Klein, "Eigenpatterns in southern California seismicity," *J. Geophys. Res.*, vol. 107, no. B12, pp. ESE 8–1–ESE 8–17, Dec. 2002, doi: 10.1029/2001jb000562. [Online]. Available: http://doi.wiley.com/10.1029/2001JB000562

[27] J. B. Rundle, A. Donnellan, G. Fox, J. P. Crutchfield, and R. Granat, "Nowcasting Earthquakes: Imaging the Earthquake Cycle in California with Machine Learning," *Life Support Biosph. Sci.*, 2021.

[28] M. J. Kane, N. Price, M. Scotch, and P. Rabinowitz, "Comparison of ARIMA and Random Forest time series models for prediction of avian influenza H5N1 outbreaks," *BMC Bioinformatics*, vol. 15, p. 276, Aug. 2014, doi: 10.1186/1471-2105-15-276. [Online]. Available: http://dx.doi.org/10.1186/1471-2105-15-276

[29] Wikipedia, "Long short-term memory." [Online]. Available: https://en.wikipedia.org/wiki/Long_short-term_memory. [Accessed: 09-Dec-2021]

[30] R. A. C. Romero, "Generative Adversarial Network for Stock Market price Prediction." [Online]. Available: http://cs230.stanford.edu/projects_fall_2019/reports/26259829.pdf. [Accessed: 09-Dec-2021]

[31] Jason Brownlee, "A Gentle Introduction to Generative Adversarial Networks (GANs)." [Online]. Available: https://machinelearningmastery.com/what-are-generative-adversarial-networks-gans/. [Accessed: 09-Dec-2021]

[32] Wikipedia, "Generative adversarial network." [Online]. Available: https://en.wikipedia.org/wiki/Generative_adversarial_network. [Accessed: 09-Dec-2021]

[33] Wikipedia, "ID3 Algorithm." [Online]. Available: https://en.wikipedia.org/wiki/ID3_algorithm

[34] Wikipedia article, "Nash–Sutcliffe model efficiency coefficient;" [Online]. Available:

https://en.wikipedia.org/wiki/Nash%E2%80%93Sutcliffe_model_efficiency_coefficient. [Accessed: 01-Dec-2020]

[35] S. Patil and M. Stieglitz, "Modelling daily streamflow at ungauged catchments: what information is necessary?," *Hydrol. Process.*, vol. 28, no. 3, pp. 1159–1169, Jan. 2014, doi: 10.1002/hyp.9660. [Online]. Available: http://doi.wiley.com/10.1002/hyp.9660

[36] D. Feng, K. Fang, and C. Shen, "Enhancing streamflow forecast and extracting insights using long-short term memory networks with data integration at continental scales," *arXiv [cs.LG]*, 18-Dec-2019 [Online]. Available: http://arxiv.org/abs/1912.08949

[37] Earthquake Hazards Program of United States Geological Survey, "USGS Search Earthquake Catalog Home Page." [Online]. Available: https://earthquake.usgs.gov/earthquakes/search/. [Accessed: 01-Dec-2020]

[38] Geoffrey Fox, "Earthquake Data Used in Study 'Earthquake Forecasting with Deep Learning.'" [Online]. Available: https://drive.google.com/drive/folders/1wz7K2R4gc78fXLNZMHcaSVfQvIpIhNPi?usp=sharing. [Accessed: 01-Dec-2020]

[39] J. B. Rundle, W. Klein, D. L. Turcotte, and B. D. Malamud, "Precursory Seismic Activation and Critical-point Phenomena," in *Microscopic and Macroscopic Simulation: Towards Predictive Modelling of the Earthquake Process*, P. Mora, M. Matsu'ura, R. Madariaga, and J.-B. Minster, Eds. Basel: Birkhäuser Basel, 2001, pp. 2165–2182 [Online]. Available: https://doi.org/10.1007/978-3-0348-7695-7_19

[40] Y. Ben-Zion and V. Lyakhovsky, "Accelerated Seismic Release and Related Aspects of Seismicity Patterns on Earthquake Faults," in *Earthquake Processes: Physical Modelling, Numerical Simulation and Data Analysis Part II*, M. Matsu'ura, P. Mora, A. Donnellan, and X.-C. Yin, Eds. Basel: Birkhäuser Basel, 2002, pp. 2385–2412 [Online]. Available: https://doi.org/10.1007/978-3-0348-8197-5_12

[41] S. Gross and J. Rundle, "A systematic test of time-to-failure analysis," *Geophys. J. Int.*, vol. 133, no. 1, pp. 57–64, Apr. 1998, doi: 10.1046/j.1365-246X.1998.1331469.x. [Online]. Available: https://academic.oup.com/gji/article-abstract/133/1/57/592353. [Accessed: 09-Dec-2021]

[42] W. I. Newman, D. L. Turcotte, and A. M. Gabrielov, "log-periodic behavior of a hierarchical failure model with applications to precursory seismic activation," *Phys. Rev. E Stat. Phys. Plasmas Fluids Relat. Interdiscip. Topics*, vol. 52, no. 5, pp. 4827–4835, Nov. 1995, doi: 10.1103/physreve.52.4827. [Online]. Available: http://dx.doi.org/10.1103/physreve.52.4827

[43] Geoffrey C. Fox, Gregor von Laszewski, Fugang Wang, Saumyadipta Pyne, "AICov: An Integrative Deep Learning Framework for COVID-19 Forecasting with Population Covariates," *J. Data Sci.*, no. Arxiv 2010.03757, p. 21, January 23, 2021, doi: 10.6339/21-JDS1007. [Online]. Available: http://dsc.soic.indiana.edu/publications/paper_covid.pdf, https://arxiv.org/abs/2010.03757

[44] B. Lim, S. O. Arik, N. Loeff, and T. Pfister, "Temporal Fusion Transformers for Interpretable Multi-horizon Time Series Forecasting," *arXiv [stat.ML]*, 19-Dec-2019 [Online]. Available: http://arxiv.org/abs/1912.09363

[45] B. Lim, S. Ö. Arık, N. Loeff, and T. Pfister, "Temporal fusion transformers for interpretable multi-horizon time series forecasting," *Int. J. Forecast.*, 2021.

[46] N. Kafritsas, "Temporal Fusion Transformer: Time Series Forecasting with Interpretability Google's state-of-the-art Transformer has it all." [Online]. Available: https://towardsdatascience.com/temporal-fusion-transformer-googles-model-for-interpretable-time-series-forecasting-5aa17beb621. [Accessed: 07-Dec-2021]

[47] Bryan Lim, Sercan Arik, Nicolas Loeff and Tomas Pfister, "Temporal Fusion Transformers for Interpretable Multi-horizon Time Series Forecasting." [Online]. Available: https://github.com/google-research/google-research/tree/master/tft. [Accessed: 07-Dec-2021]

[48] "TFT For PyTorch." [Online]. Available: https://catalog.ngc.nvidia.com/orgs/nvidia/resources/tft_for_pytorch. [Accessed: 07-Dec-2021]

[49] B. Feng and G. C. Fox, "Spatiotemporal Pattern Mining for Nowcasting Extreme Earthquakes in Southern California," in *Proceedings of 2021 IEEE 17th International Conference on eScience*, Innsbruck, Austria, 2021, pp. 99–107.

[50] B. Feng and G. C. Fox, "TSEQPREDICTOR: Spatiotemporal Extreme Earthquakes Forecasting for Southern California," *arXiv preprint arXiv:2012.14336*, 2020.

[51] Fox, Geoffrey, "Study of Earthquakes with Deep Learning (Earthquakes for Real); Lectures in class on AI First Engineering," 22-Apr-2021. [Online]. Available: https://docs.google.com/presentation/d/1ykYnX0uvxPE-M-c-Tau8irU3IqYuvj8Ws8iUqd5RCxQ/edit?usp=sharing. [Accessed: 30-Nov-2021]

[52] Geoffrey Fox, "Deep Learning Based Time Evolution." [Online]. Available: http://dsc.soic.indiana.edu/publications/Summary-DeepLearningBasedTimeEvolution.pdf. [Accessed: 08-Jun-2020]

[53] G. Fox, "FFFFWNPF-EARTHQB-LSTMFullProps2 Google Colab for LSTM Forecast." [Online]. Available: https://colab.research.google.com/drive/16DjDXv8wjzNm7GABNMCGiE-Q0gFAlNHz?usp=sharing. [Accessed: 07-Dec-2021]

[54] G. Fox, "FFFFWNPFEARTHQ-newTFTv29 Google Colab for TFT Forecast." [Online]. Available: https://colab.research.google.com/drive/12zEv08wvwRhQEwYWy641j9dLSDskxooG?usp=sharing. [Accessed: 07-Dec-2021]

[55] G. Fox, "Y8FFFFWNPF-EARTHQDGX-Transformer1 DGX Jupyter Notebook for Science Transformer Forecast," *https://drive.google.com/file/d/1Mj6RgS_AqGdzQoSeGfKxsPqTQ8OJGJ84/view?usp=sharing*. .

[56] A. Galassi, M. Lippi, and P. Torroni, "Attention in Natural Language Processing," *arXiv [cs.CL]*, 04-Feb-2019 [Online]. Available: http://arxiv.org/abs/1902.02181

[57] D. A. Kaji *et al.*, "An attention based deep learning model of clinical events in the intensive care unit," *PLoS One*, vol. 14, no. 2, p. e0211057, Feb. 2019, doi: 10.1371/journal.pone.0211057. [Online]. Available: http://dx.doi.org/10.1371/journal.pone.0211057

[58] T. Gangopadhyay, S. Y. Tan, Z. Jiang, R. Meng, and S. Sarkar, "Spatiotemporal Attention for Multivariate Time Series Prediction and Interpretation," *arXiv [cs.LG]*, 11-Aug-2020 [Online]. Available: http://arxiv.org/abs/2008.04882

[59] N. Xu, Y. Shen, and Y. Zhu, "Attention-Based Hierarchical Recurrent Neural Network for Phenotype Classification," in *Advances in Knowledge Discovery and Data Mining*, 2019, pp. 465–476, doi: 10.1007/978-3-030-16148-4_36 [Online]. Available: http://dx.doi.org/10.1007/978-3-030-16148-4_36

[60] R. S. Kodialam, R. Boiarsky, and D. Sontag, "Deep Contextual Clinical Prediction with Reverse Distillation," *arXiv [cs.LG]*, 10-Jul-2020 [Online]. Available: http://arxiv.org/abs/2007.05611

[61] J. Gao *et al.*, "CAMP: Co-Attention Memory Networks for Diagnosis Prediction in Healthcare," in *2019 IEEE International Conference on Data Mining (ICDM)*, 2019, pp. 1036–1041, doi: 10.1109/ICDM.2019.00120 [Online]. Available: http://dx.doi.org/10.1109/ICDM.2019.00120

[62] R. Sen, H.-F. Yu, and I. Dhillon, "Think Globally, Act Locally: A Deep Neural Network Approach to High-Dimensional Time Series Forecasting," *arXiv [stat.ML]*, 09-May-2019 [Online]. Available: http://arxiv.org/abs/1905.03806

[63] H. Song, D. Rajan, J. J. Thiagarajan, and A. Spanias, "Attend and diagnose: Clinical time

series analysis using attention models," in *Thirty-second AAAI conference on artificial intelligence*, 2018 [Online]. Available: https://www.aaai.org/ocs/index.php/AAAI/AAAI18/paper/viewPaper/16325
[64] A. Vaswani *et al.*, "Attention is All you Need," in *Advances in Neural Information Processing Systems*, 2017, vol. 30, pp. 5998–6008 [Online]. Available: https://proceedings.neurips.cc/paper/2017/file/3f5ee243547dee91fbd053c1c4a845aa-Paper.pdf
[65] A. Vaswani *et al.*, "Attention Is All You Need," *arXiv [cs.CL]*, 12-Jun-2017 [Online]. Available: http://arxiv.org/abs/1706.03762
[66] A. Zeyer, P. Bahar, K. Irie, R. Schlüter, and H. Ney, "A Comparison of Transformer and LSTM Encoder Decoder Models for ASR," in *2019 IEEE Automatic Speech Recognition and Understanding Workshop (ASRU)*, 2019, pp. 8–15, doi: 10.1109/ASRU46091.2019.9004025 [Online]. Available: http://dx.doi.org/10.1109/ASRU46091.2019.9004025
[67] Z. Zeng *et al.*, "Leveraging Text Data Using Hybrid Transformer-LSTM Based End-to-End ASR in Transfer Learning," *arXiv [eess.AS]*, 21-May-2020 [Online]. Available: http://arxiv.org/abs/2005.10407
[68] X. Huang, G. C. Fox, S. Serebryakov, A. Mohan, P. Morkisz, and D. Dutta, "Benchmarking Deep Learning for Time Series: Challenges and Directions," in *2019 IEEE International Conference on Big Data (Big Data)*, 2019, pp. 5679–5682, doi: 10.1109/BigData47090.2019.9005496 [Online]. Available: http://dx.doi.org/10.1109/BigData47090.2019.9005496
[69] J. Kates-Harbeck, A. Svyatkovskiy, and W. Tang, "Predicting disruptive instabilities in controlled fusion plasmas through deep learning," *Nature*, vol. 568, no. 7753, pp. 526–531, Apr. 2019, doi: 10.1038/s41586-019-1116-4. [Online]. Available: https://doi.org/10.1038/s41586-019-1116-4
[70] Geoffrey Fox, Tony Hey, Jeyan Thiyagalingam, "Science Data working Group of MLCommons." [Online]. Available: https://mlcommons.org/en/groups/research-science/. [Accessed: 03-Dec-2020]
[71] "MLCommons Homepage: Machine learning innovation to benefit everyone." [Online]. Available: https://mlcommons.org/en/. [Accessed: 07-Dec-2021]